\documentclass[twocolumn, aps, pra, superscriptaddress, floatfix]{revtex4}
\usepackage[T1]{fontenc}
\usepackage[latin9]{inputenc}
\usepackage{color}
\usepackage{bm}
\usepackage{xcolor}
\usepackage{graphics}
\usepackage[sort&compress]{natbib}
\usepackage{amsmath,amsthm,verbatim,amssymb,amsfonts,amscd}

\usepackage{braket}
\usepackage{psfrag}
\usepackage{tikz}
\usepackage[normalem]{ulem}
\usepackage{qcircuit}
\usepackage{subcaption}
\usepackage{accents}
\usepackage{cleveref}
\usepackage{soul}
\usepackage{multirow}
\usepackage{tkz-euclide}

\usepackage{pgfplots}
\usetikzlibrary{pgfplots.groupplots}
\usepackage{physics}
\usepackage{graphics}
\usepackage{float}
\usepackage{caption}
\theoremstyle{definition}
\newtheorem{theorem}{Theorem}
\newtheorem{remark}{Remark}
\newtheorem{lemma}{Lemma}

\usepackage{listings}
\usepackage{mathtools}
\DeclarePairedDelimiter\ceil{\lceil}{\rceil}
\DeclarePairedDelimiter\floor{\lfloor}{\rfloor}

\newcommand{\defeq}{\vcentcolon=}
\newcommand{\eqdef}{=\vcentcolon}

\usepackage[ruled,vlined,linesnumbered]{algorithm2e}
\usepackage[noend]{algpseudocode}

 \newlength\figureheight
\newlength\figurewidth

\begin{document}

\title{Iterative Quantum Amplitude Estimation}

\author{Dmitry Grinko}%
\affiliation{IBM Quantum, IBM Research -- Zurich}
\affiliation{ETH Zurich}
\affiliation{University of Geneva}

\author{Julien Gacon}
\affiliation{IBM Quantum, IBM Research -- Zurich}
\affiliation{ETH Zurich}

\author{Christa Zoufal}
\affiliation{IBM Quantum, IBM Research -- Zurich}
\affiliation{ETH Zurich}

\author{Stefan Woerner}
\email{wor@zurich.ibm.com}
\affiliation{IBM Quantum, IBM Research -- Zurich}
		
\date{\today}

\begin{abstract}
We introduce a variant of \emph{Quantum Amplitude Estimation (QAE)}, called \emph{Iterative QAE} (IQAE), which does not rely on \emph{Quantum Phase Estimation} (QPE) but is only based on \emph{Grover's Algorithm}, which reduces the required number of qubits and gates.
We provide a rigorous analysis of IQAE and prove that it achieves a quadratic speedup up to a double-logarithmic factor compared to classical Monte Carlo simulation with provably small constant overhead.
Furthermore, we show with an empirical study that our algorithm outperforms other known QAE variants without QPE, some even by orders of magnitude, i.e., our algorithm requires significantly fewer samples to achieve the same estimation accuracy and confidence level.
\end{abstract}

\maketitle

\section{Introduction}

\emph{Quantum Amplitude Estimation} (QAE) \cite{brassard} is a fundamental quantum algorithm with the potential to achieve a quadratic speedup for many applications that are classically solved through Monte Carlo (MC) simulation.
It has been shown that we can leverage QAE in the financial service sector, e.g., for risk analysis \cite{wor, Egger2019} or option pricing \cite{Rebentrost2018, Stamatopoulos2019, Zoufal2019}, and also for generic tasks such as numerical integration \cite{Montanaro2017}.
While the estimation error bound of classical MC simulation scales as $\mathcal{O}(1/\sqrt{M})$, where $M$ denotes the number of (classical) samples, QAE achieves a scaling of $\mathcal{O}(1/M)$ for $M$ (quantum) samples, indicating the aforementioned quadratic speedup.

The canonical version of QAE is a combination of \emph{Quantum Phase Estimation} (QPE) \cite{nielsen} and \emph{Grover's Algorithm}.
Since other QPE-based algorithms are believed to achieve exponential speedup, most prominently \emph{Shor's Algorithm} for factoring \cite{Shor1997}, it has been speculated as to whether QAE can be simplified such that it uses only Grover iterations without a QPE-dependency.
Removing the QPE-dependency would help to reduce the resource requirements of QAE in terms of qubits and circuit depth and lower the bar for practial applications of QAE.

Recently, several approaches have been proposed in this direction.
In \cite{Suzuki2019} the authors show how to replace QPE by a set of Grover iterations combined with a Maximum Likelihood Estimation (MLE), in the following called \emph{Maximum Likelihood Amplitude Estimation} (MLAE).
In \cite{Wie2019}, QPE is replaced by the Hadamard test, analog to \emph{Kitaev's Iterative QPE} \cite{Kitaev1995, kitaev2002classical} and similar approaches \cite{svore2013faster,atia2017fast}.

Both \cite{Suzuki2019} and \cite{Wie2019} propose potential simplifications of QAE, but do not provide rigorous proofs of the correctness of the proposed algorithms.
In \cite{Wie2019}, it is not even clear how to control the accuracy of the algorithm other than possibly increasing the number of measurements of the evolving quantum circuits.
Thus, the potential quantum advantage is difficult to compare and we will not discuss it in the remainder of this paper.

In \cite{Aaronson2019}, another variant of QAE was proposed.
There, for the first time, it was rigorously proven that QAE without QPE can achieve a quadratic speedup over classical MC simulation.
Following \cite{Aaronson2019}, we call this algorithm \emph{QAE, Simplified} (QAES).
Although this algorithm achieves the desired asymptotic complexity exactly (i.e. without logarithmic factors), the involved constants are very large, and likely to render this algorithm impractical unless further optimized -- as shown later in this manuscript.

In the following, we propose a new version of QAE -- called \emph{Iterative QAE} (IQAE) -- that achieves better results than all other tested algorithms.
It provably has the desired asymptotic behavior up to a multiplicative $\log(2/\alpha \log_2(\pi/4\epsilon))$ factor, where $\epsilon > 0$ denotes the target accuracy, and $1-\alpha$ the resulting confidence level.

Like in \cite{Aaronson2019}, our algorithm requires iterative queries to the quantum computer to achieve the quadratic speedup and cannot be parallelized.
Only MLAE allows the parallel execution of the different queries as the estimate is derived via classical MLE applied to the results of all queries.
Although parallelization is a nice feature, the potential speedup is limited.
Assuming the length of the queries is doubled in each iteration (like for canonical QAE and MLAE) the speedup is at most a factor of two, since the computationally most expensive query dominates all the others.

With MLAE, QAES, and IQAE we have three promising variants of QAE that do not require QPE and it is of general interest to empirically compare their performance.
Of similar interest is the question whether the the canonical QAE with QPE -- while being (quantum) computationally more expensive -- might lead to some performance benefits.
To be able to better compare the performance of canonical QAE with MLAE, QAES, and IQAE, we extend QAE by a classical MLE postprocessing based on the observed results.
This improves the results without additional queries to the quantum computer and allows us to derive proper confidence intervals.

The remainder of this paper is organized as follows.
Sec.~\ref{sec:qae} introduces QAE in its canonical form, its considered variants, as well as the proposed MLE postprocessing.
In Sec.~\ref{sec:iqae}, we introduce IQAE and provide the corresponding theoretical results.
Empirical results, comparing the performance of the different algorithms on various test cases, are reported in Sec.~\ref{sec:results} and illustrate the efficiency of our new algorithm.
To conclude, we discuss our results and open questions in Sec.~\ref{sec:conclusion}.

\section{Quantum Amplitude Estimation} \label{sec:qae}

QAE was first introduced in \cite{brassard} and assumes the problem of interest is given by an operator $\mathcal{A}$ acting on $n+1$ qubits such that
\begin{eqnarray}
\mathcal{A}\ket{0}_n\ket{0} = \sqrt{1-a} \ket{\psi_0}_n\ket{0} + \sqrt{a}\ket{\psi_1}_n\ket{1},
\end{eqnarray}
where $a \in [0, 1]$ is the unknown, and $\ket{\psi_0}_n$ and $\ket{\psi_1}_n$ are two normalized states, not necessarily orthogonal.
QAE allows to estimate $a$ with high probability such that the estimation error scales as $\mathcal{O}(1/M)$, where $M$ corresponds to the number of applications of $\mathcal{A}$.
To this extent, an operator $\mathcal{Q} = \mathcal{A} \mathcal{S}_0 \mathcal{A}^{\dagger} \mathcal{S}_{\psi_0}$ is defined where $\mathcal{S}_{\psi_0} = \mathbb{I} - 2 \ket{\psi_0}\bra{\psi_0} \otimes \ket{0}\bra{0}$ and $\mathcal{S}_{0} = \mathbb{I} - 2 \ket{0}_{n+1} \bra{0}_{n+1}$ as introduced in \cite{brassard}.
In the following, we denote applications of  $\mathcal{Q}$ as \emph{quantum samples} or \emph{oracle queries}.

\begin{figure}[hbtp]
\centering
\includegraphics[width=0.45\textwidth]{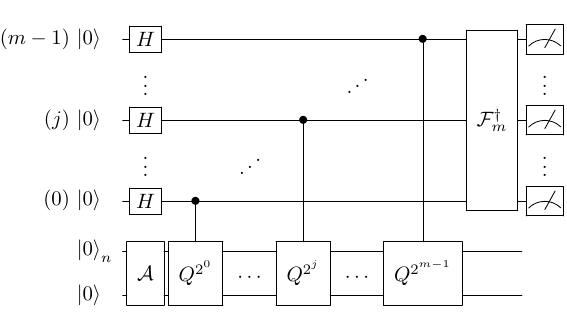}
\caption{QAE circuit with $m$ ancilla qubits and $n+1$ state qubits.}
\label{fig:qae}
\end{figure}

The canonical QAE follows the form of QPE: 
it uses $m$ ancilla qubits -- initialized in equal superposition -- to represent the final result, it defines the number of quantum samples as $M = 2^m$ and applies geometrically increasing powers of $\mathcal{Q}$ controlled by the ancillas. Eventually, it performs a QFT on the ancilla qubits before they are measured, as illustrated in Fig.~\ref{fig:qae}.
Subsequently, the measured integer $y \in \{0, \ldots, M-1\}$ is mapped to an angle $\tilde{\theta}_a = y\pi/M$.
Thereafter, the resulting estimate of $a$ is defined as $\tilde{a} = \sin^2(\tilde{\theta}_a)$.
Then, with a probability of at least $8/\pi^2 \approx 81\%$, the estimate $\tilde{a}$ satisfies  
\begin{eqnarray}
|a - \tilde{a}| \leq \frac{2\pi\sqrt{a(1-a)}}{M} + \frac{\pi^2}{M^2},
\end{eqnarray}
which implies the quadratic speedup over a classical MC simulation, i.e., the estimation error $\epsilon = \mathcal{O}(1/M)$.
The success probability can quickly be boosted to close to $100\%$ by repeating this multiple times and using the median estimate \cite{wor}.
These estimates $\tilde{a}$ are restricted to the grid $\left\{\sin^2\left(y\pi/M\right) : y = 0, \ldots, M  / 2 \right\}$ through the possible measurement outcomes of $y$.

Alternatively, and similarly to MLAE, it is possible to apply MLE to the observations for $y$.
For a given $\theta_a$, the probability of observing $\ket{y}$ when measuring the ancilla qubits is derived in \cite{brassard} and given by 
\begin{eqnarray}
\mathbb{P}[\ket{y}] = \frac{\sin^2(M \Delta \pi)}{M^2 \sin^2(\Delta \pi)}, \label{eq:qae_sample_probability}
\end{eqnarray}
where $\Delta$ is the minimal distance on the unit circle between the angles $\theta_a$ and $\pi \tilde{y}/M$, and $\tilde{y} = y$ if $y \leq M/2$ and $\tilde{y} = M/2 - y$ otherwise.
Given a set of $y$-measurements, this can be leveraged in an MLE to get an estimate of $\theta_a$ that is not restricted to grid points.
Furthermore, it allows to use the likelihood ratio to derive confidence intervals \cite{koch_1999_parameter}.
This is discussed in more detail in Appendix \ref{sec:canonical_qae_with_mle}.
In our tests, the likelihood ratio confidence intervals were always more reliable than other possible approaches, such as the (observed) Fisher  information.
Thus, in the following, we will use the term QAE for the canonical QAE with the application of MLE to the $y$ measurements to derive an improved estimate and confidence intervals based on the likelihood ratio.

All variants of QAE without QPE -- including ours -- are based on the fact that 
\begin{eqnarray}
\mathcal{Q}^{k} \mathcal{A} \ket{0}_n\ket{0} &=&
\cos((2k+1)\theta_a)\ket{\psi_0}_n\ket{0} + \nonumber \\
&& \sin((2k+1)\theta_a)\ket{\psi_1}_n\ket{1},
\end{eqnarray}
where $\theta_a$ is defined as $a = \sin^2(\theta_a)$.
In other words, the probability of measuring $\ket{1}$ in the last qubit is given by
\begin{equation}
\mathbb{P}[\ket{1}] = \sin^2((2k+1)\theta_a).
\end{equation}
The algorithms mainly differ in how they derive the different values for the powers $k$ of $\mathcal{Q}$ and how they combine the results into a final estimate of $a$.

MLAE first approximates $\mathbb{P}[\ket{1}]$ for $k = 2^j$ and $j=0, 1, 2, \ldots, m-1$, for a given $m$, using $N_{\text{shots}}$ measurements from a quantum computer for each $j$, i.e., in total, $\mathcal{Q}$ is applied $N_{\text{shots}}(M-1)$ times, where $M = 2^m$.
It has been shown in \cite{Suzuki2019} that the corresponding Fisher information scales as $\mathcal{O}(N_{\text{shots}} M^2)$, which implies a lower bound of the estimation error scaling as $\Omega(1/(\sqrt{N_{\text{shots}}}M))$.
Crucially, \cite{Suzuki2019} does not provide an upper bound for the estimation error.
Confidence intervals can be derived from the measurements using, e.g., the likelihood ratio approach, see Appendix \ref{sec:canonical_qae_with_mle}.

In contrast to MLAE, QAES requires the different powers of $\mathcal{Q}$ to be evaluated iteratively and cannot be parallelized.
It iteratively adapts the powers of $\mathcal{Q}$ to successively improve the estimate and carefully determines the next power of $\mathcal{Q}$.
However, instead of a lower bound, a rigorous error upper bound is provided.
QAES achieves the optimal asymptotic query complexity $\mathcal{O}(\log(1/\alpha)/\epsilon)$, where $\alpha>0$ denotes the probability of failure.
In contrast to the other algorithms considered, QAES provides a bound on the relative estimation error.
Although the algorithm achieves the desired asymptotic scaling exactly, the constants involved are very large -- likely too large for practical applications unless they can be further reduced.

In the following, we introduce a new variant of QAE without QPE.
As for QAES, we provide a rigorous performance proof.
Although our algorithm only achieves the quadratic speedup up to a multiplicative factor $\log(2/\alpha \log_2(\pi/4\epsilon))$, the constants involved are orders of magnitude smaller than for QAES. Moreover, in practice this doubly logarithmic factor is small for any reasonable target accuracy $\epsilon$ and confidence level $1-\alpha$, as we will show in \ref{sec:results}.

\section{Iterative Quantum Amplitude Estimation} \label{sec:iqae}

IQAE leverages similar ideas as \cite{Suzuki2019, Wie2019, Aaronson2019} but combines them in a different way, which results in a more efficient algorithm while still allowing for a rigorous upper bound on the estimation error and computational complexity.
As mentioned before, we use the quantum computer to approximate $\mathbb{P}[\ket{1}] = \sin^2((2k+1)\theta_a)$ for the last qubit in $\mathcal{Q}^k \mathcal{A} \ket{0}_n\ket{0}$ for different powers $k$.
In the following, we outline the rationale behind IQAE, which is formally given in Alg.~\ref{alg:iqae}. The main sub-routine \textsc{FindNextK} is outlined in Alg.~\ref{alg:find_k}.

\begin{figure}[h] 
\includegraphics[width=0.5\textwidth]{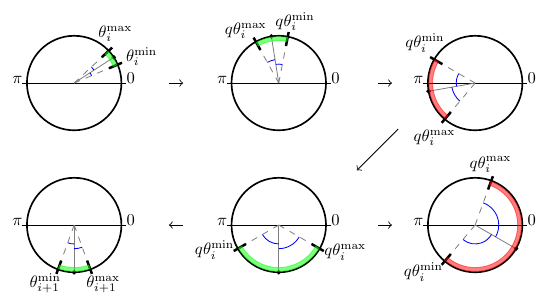}
\caption{
\label{fig:findK}
\textsc{FindNextK}: Given an initial interval $[\theta_l,\theta_u]$, $k_i$, and $K_i = 4k_i + 2$, \textsc{FindNextK} determines the largest feasible $k$ with $K = 4k+2 \geq 2K_i$ such that the scaled interval $[K\theta_l,K\theta_u]_{\text{mod} 2\pi}$ lies either in the upper or in the lower half-plane, and returns $k$ if it exists and $k_i$ otherwise.
The top left circle represents our initial knowledge about ${K_i\theta_a}$, while other circles represent extrapolations for different values of $q = K/K_i$.
The top middle picture represents a valid $q$, the top right circle represents an invalid $q$, and so on. 
Note that the bottom right circle violates the condition $q \cdot \abs{\theta_i^{\text{max}} - \theta_i^{\text{min}}} \leq \pi$, i.e., the interval is too wide and cannot lie in a single half-plane.
The output of \textsc{FindNextK} is the middle bottom circle, and the left bottom figure shows the improved result in the next iteration after additional measurements.
}
\end{figure}

Suppose a confidence interval $[\theta_l, \theta_u] \subseteq [0, \pi/2]$ for $\theta_a$ and a power $k$ of $\mathcal{Q}$ as well as an estimate for $\sin^2((2k+1)\theta_a)$.
Through exploiting the trigonometric identity $\sin^2(x) = (1 - \cos(2x))/2$, we can translate our estimates for $\sin^2((2k+1)\theta_a)$ into estimates for $\cos((4k + 2)\theta_a)$.
However, unlike in Kitaev's Iterative QPE, we cannot estimate the sine, and the cosine alone is only invertible without ambiguity if we know the argument is restricted to either $[0, \pi]$ or $[\pi, 2\pi]$, i.e., the upper or lower half-plane.
Thus, we want to find the largest $k$ such that the scaled interval $[(4k+2)\theta_l, (4k+2)\theta_u]_{\text{mod } 2\pi}$ is fully contained either in $[0, \pi]$ or $[\pi, 2\pi]$.
If this is given, we can invert $\cos((4k + 2)\theta_a)$ and improve our estimate for $\theta_a$ with high confidence.
This implies an upper bound of $k$, and the heart of the algorithm is the procedure used to find the next $k$ given $[\theta_l, \theta_u]$, which is formally introduced in Alg.~\ref{alg:find_k} and illustrated in Fig.~\ref{fig:findK}.
In the following theorem, we provide convergence results for IQAE that imply the aforementioned quadratic speedup.
The respective proof is given in Appendix \ref{sec:proof_of_convergence}.

More intuitively, the heart of our algorithm - the sub-routine \textsc{FindNextK} - allows us to maximize Fisher Information $\mathcal{I}$ on a given iteration in a greedy fashion. The way to see this is to notice, that any summand of $\mathcal{I}$ is proportional to $N_{\mathrm{shots}}K^2$, where $K\defeq4k+2$ \cite{Suzuki2019}.

\begin{algorithm}
\caption{\label{alg:iqae} Iterative Quantum Amplitude Estimation}
\DontPrintSemicolon
\SetKwFunction{FFind}{FindNextK}
\SetKwFunction{FMain}{IQAE}
\SetKwProg{Fn}{Function}{:}{}
\Fn{\FMain{$\epsilon$, $\alpha$, $N_{\mathrm{shots}}$, $\mathrm{ci}$}}
{
    \tcp{ci is a chosen confidence interval method, which can be either Clopper-Pearson \cite{ClopperPearson1934,scholz2008confidence} or Chernoff-Hoeffding \cite{hoeffding1963probability}}
    $i = 0$ \tcp{initialize iteration count}
	$k_i = 0$ \tcp{initialize power of $\mathcal{Q}$}
	$\mathrm{up}_i = \mathrm{True}$ \tcp{keeps track of the half-plane}
	$[\theta_l, \theta_u] = [0, \pi/2]$ \tcp{initialize conf.~interval}
	$T = \ceil{\log_2(\pi / 8\epsilon)}$ \tcp{max.~number of rounds}
	calculate $L_{\text{max}}$ according to (\ref{def:LmaxCHgen})-(\ref{def:LmaxCPgen}) \tcp{max.~error on every iteration; depends on $\epsilon$, $\alpha$, $N_{\mathrm{shots}}$ and choice of confidence interval}
    \While{$\theta_u-\theta_l > 2\epsilon$}
    {
        $i = i+1$\;    
        $k_i, \mathrm{up}_i$ = \FFind{$k_{i-1}, \theta_l, \theta_u, \mathrm{up}_{i-1}$}\; 

 		set $K_i = 4k_i+2$ \;    
 		
 		\If{$K_i > \ceil{L_{\mathrm{max}} /\epsilon}$}{
 		$N = \ceil{ N_{\mathrm{shots}} L_{\text{max}} /\epsilon/K_i/10 }$
 		\tcp{No-overshooting condition}
 		}\Else{ 
 		$N = N_{\mathrm{shots}}$
 		}
        
        approximate $a_i = \mathbb{P}[\ket{1}]$ for the last qubit of $\mathcal{Q}^{k_i} \mathcal{A} \ket{0}_n\ket{0}$ by measuring $N$ times\;
        \If{$k_i = k_{i-1}$}
        {
            combine the results of all iterations $j \leq i$ with $k_j = k_i$ into a single results, effectively increasing the number of shots
        }
        \If{$\text{\normalfont ci = "Chernoff-Hoeffding"}$}{
        $\epsilon_{a_i} = \sqrt{\frac{1}{2N}\log(\frac{2T}{\alpha})}$ \\
        $a_i^{\text{max}} = \min(1, a_i + \epsilon_{a_i})$ \\
        $a_i^{\text{min}} = \max(0, a_i - \epsilon_{a_i})$
        }
        \If{$\text{\normalfont ci = "Clopper-Pearson"}$}{
        $a_i^{\text{max}} = I^{-1}(\frac{\alpha}{2T} ; Na_i , N(1 - a_i) + 1)$ \\
        $a_i^{\text{min}} = I^{-1}(1 - \frac{\alpha}{2T} ; Na_i + 1 , N(1 - a_i))$ \tcp{see equations (\ref{eq:cp_min_def}-\ref{eq:clopper-pearson-i})}}
        calculate the confidence interval $[\theta_i^{\text{min}}, \theta_i^{\text{max}}]$ for $\{K_i \theta_a\}_{\text{mod } 2\pi}$ from $[a^{\text{min}}_i, a^{\text{max}}_i]$ and boolean flag $\mathrm{up}_i$ by inverting $a = (1 - \cos(K_i\theta))/2$\;
    
        $\theta_l =  \frac{\floor{K_i\theta_l}_{\text{mod } 2\pi}+\theta_i^{\text{min}}}{K_i}$\;
        $\theta_u =  \frac{\floor{K_i\theta_u}_{\text{mod } 2\pi}+\theta_i^{\text{max}}}{K_i}$\;
    }
    $[a_l, a_u] = [\sin^2(\theta_l), \sin^2(\theta_u)]$\;
    \KwRet $[a_l, a_u]$\;
}
\end{algorithm}

\newpage
\begin{algorithm}
\caption{Procedure for finding $k_{i+1}$ \label{alg:find_k}}
\DontPrintSemicolon
\SetKwFunction{FFind}{FindNextK}
\SetKwProg{Fn}{Function}{:}{}
\Fn{\FFind{$k_i,  \theta_l, \theta_u, \mathrm{up}_i, r = 2$}}
{
    $K_i = 4k_i +2$ \tcp{current $\theta$-factor}
  	$\theta_i^{\text{min}} = K_i \theta_l$ \tcp{lower bound for scaled $\theta$}
    $\theta_i^{\text{max}} = K_i \theta_u$ \tcp{upper bound for scaled $\theta$}
     $K_{\text{max}} = \floor{\frac{\pi}{\theta_u-\theta_l}} $\tcp{set an upper bound for $\theta$-factor}
	$K = K_{\text{max}} - (K_{\text{max}}-2)_{\text{ mod } 4}$ \tcp{largest potential candidate of the form $4k+2$}
    \While{$K \geq r K_i$}
    {
        $q = K/K_i$ \tcp{factor to scale $[\theta_i^{\text{min}}, \theta_i^{\text{max}}]$}
        \If{$ \{ q \cdot \theta_i^{\text{max}} \}_{\text{mod }2\pi} \leq \pi \text{ \textbf{and} } \{ q \cdot \theta_i^{\text{min}} \}_{\text{mod }2\pi} \leq \pi$}
        { 
            \tcp{$[\theta_{i+1}^{\text{min}}, \theta_{i+1}^{\text{max}}]$ is in upper half-plane}
            $K_{i+1} = K$\;
            $\mathrm{up}_{i+1} = \text{True}$\;
            $k_{i+1} = (K_{i+1}-2)/4$\;
            \KwRet $(k_{i+1},\mathrm{up}_{i+1})$
        }
        \If{$ \{ q \cdot \theta_i^{\text{max}} \}_{\text{mod }2\pi} \geq \pi \text{ \textbf{and} } \{ q \cdot \theta_i^{\text{min}} \}_{\text{mod }2\pi} \geq \pi$} 
        {
            \tcp{$[\theta_{i+1}^{\text{min}}, \theta_{i+1}^{\text{max}}]$ is in lower half-plane}
            $K_{i+1} = K$\; 
            $\mathrm{up}_{i+1} = \text{False}$\;
            $k_{i+1} = (K_{i+1}-2)/4$\;
            \KwRet $(k_{i+1},\mathrm{up}_{i+1})$
        }
        $K = K - 4$
    }
    \KwRet $(k_i,\mathrm{up_i})$ \tcp{return old value}
}
\end{algorithm}

\begin{theorem}[Correctness of IQAE]\label{thm:proof_of_convergence}
Suppose a confidence level $1 - \alpha \in (0, 1)$, a target accuracy $\epsilon > 0$, and a number of shots $N_{\text{shots}} \in \{1, ..., N_{\text{max}}(\epsilon, \alpha)\}$, where 
\begin{eqnarray}N_{\text{max}}(\epsilon, \alpha) = \frac{32}{(1-2\sin(\pi/14))^2} \log(\frac{2}{\alpha} \log_2 \left( \frac{\pi}{4\epsilon} \right)).
\label{eq:n_max_ch}
\end{eqnarray}
In this case, IQAE (Alg.~\ref{alg:iqae}) terminates after a maximum number of $\ceil{\log_2(\pi/8\epsilon)}$ rounds, where we define one round as a set of iterations with the same $k_i$, and each round consists of at most $N_{\text{max}}(\epsilon, \alpha) / N_{\text{shots}}$ iterations.
IQAE computes $[\theta_l, \theta_u]$ with $\theta_u - \theta_l \leq 2\epsilon$ and
\begin{eqnarray}
\mathbb{P}[\theta_a \notin [\theta_l, \theta_u]] \leq \alpha,
\end{eqnarray}
and returns $[a_l, a_u]$ with $a_u - a_l \leq 2\epsilon$ and
\begin{eqnarray}
\mathbb{P}[a \notin [a_l, a_u]] \leq \alpha.
\end{eqnarray}
Thus, $\tilde{a} = (a_l + a_u)/2$ leads to an estimate for $a$ with $|a - \tilde{a}| \leq \epsilon$ with a confidence of $1-\alpha$.

Furthermore, for the total number of $\mathcal{Q}$-applications, $N_{\text{oracle}}$, it holds that
\begin{eqnarray}
N_{\text{oracle}} 
&<& \frac{50}{\epsilon} \log(\frac{2}{\alpha}\log_2 \left (\frac{\pi}{4\epsilon}  \right)).
\end{eqnarray}
\end{theorem}

Note that the maximum number of applications of $\mathcal{Q}$ given in Thm.~\ref{thm:proof_of_convergence} is a loose upper bound since the proof uses Chernoff-Hoeffding bound to estimate sufficiently narrow intermediate confidence intervals in Alg.~\ref{alg:iqae}. 
Using more accurate techniques instead, such as Clopper-Pearson's confidence interval for Bernoulli distributions \cite{ClopperPearson1934}, can lower the constant overhead in $N_{\text{oracle}}$ by a factor of 3 (see Appendix \ref{sec:proof_of_convergence_CP}) but is more complex to analyze analytically. 
In Sec.~\ref{sec:results}, we demonstrate how this two approaches perform empirically.

In Alg.~\ref{alg:find_k}, we require that $K_{i+1} / K_i \geq  r = 2$, otherwise we continue with $K_i$. The choice of the lower bound $r$ is optimal in the proof, i.e. it gives us the lowest coefficient for the upper bound (see Appendix~\ref{thm:proof_of_convergence}).
Moreover, the chosen lower bound was working very well  in practice.

In Alg.~\ref{alg:iqae} we imposed the "no-overshooting" condition in order to ensure, that we do not make unnecessary measurement shots at last iterations of the algorithm. This condition also allows us to keep constants small in the proof (see condition (\ref{eq:Kt})). It utilizes a quantity $L_{\text{max}}$ - the maximum possible error, which could be returned on a given iteration using $N_{\text{shots}}$ measurements.
It is calculated before the start of the algorithm for chosen $\epsilon$, $\alpha$ and number of shots $N_{\text{shots}}$. It also depends on the type of chosen confidence interval.
For Chernoff-Hoeffding one can write a direct analytical expression:
\begin{equation}
L_{\text{max}}(N_{\text{shots}},\epsilon,\alpha) \defeq \arcsin \left( \frac{2}{N_{\text{shots}}}\log(\frac{2T(\epsilon)}{\alpha}) \right)^{1/4}, \label{def:LmaxCHgen}
\end{equation}
which is derived from equations (\ref{eq:n_max_bound}) and (\ref{def:LmaxCH}) from Appendix \ref{sec:proof_of_convergence}.
For Clopper-Pearson you can only calculate it numerically:
\begin{equation}
L_{\text{max}}(N_{\text{shots}},\epsilon,\alpha) \defeq \max_{\theta} h_{N_{\text{shots}},\epsilon,\alpha}(\theta),
\label{def:LmaxCPgen}
\end{equation}
where function $h$ is defined in Appendix \ref{sec:proof_of_convergence_CP}. It is derived by analogy with formula (\ref{def:LmaxCP}), where instead of $N_{\text{max}}(\epsilon,\alpha)$ one should use $N_{\text{shots}}$.

Thm.~\ref{thm:proof_of_convergence} provides a bound on the query complexity, i.e., the total number of oracle calls with respect to the target accuracy.
However, it is important to note that the computational complexity, i.e., the overall number of operations, including classical steps such as all applications of \textsc{FindNextK} and computing the intermediate confidence intervals, scales in exactly the same way.

\section{Results} \label{sec:results}

In this section, we empirically compare IQAE, MLAE, QAES, QAE, and classical MC with each other and determine the total number of oracle queries necessary to achieve a particular accuracy.
We are only interested in measuring the last qubit of $\mathcal{Q}^k \mathcal{A} \ket{0}_n\ket{0}$ for different powers $k$, and we know that $\mathbb{P}[\ket{1}] = \sin^2((2k+1)\theta_a)$.
Thus, for a given $\theta_a$ and $k$, we can consider a Bernoulli distribution with corresponding success probability or a single-qubit $R_y$-rotation with angle $2(2k+1)\theta_a$ to generate the required samples.
All algorithms mentioned in this paper are implemented and tested using Qiskit \cite{qiskit} in order to be run on simulators or real quantum hardware, e.g., as provided via  the \emph{IBM Quantum Experience}.

For IQAE and MC, we compute the (intermediate) confidence intervals based both on Chernoff-Hoeffding \cite{hoeffding1963probability} and on Clopper-Pearson \cite{ClopperPearson1934}.
For QAE and MLAE, we use the likelihood ratio \cite{koch_1999_parameter}, cf.~Appendix \ref{sec:canonical_qae_with_mle}.
For QAES, we report the outputted accuracy of the algorithm.

\begin{figure*}[hbtp]
\centering
\includegraphics[width=\textwidth]{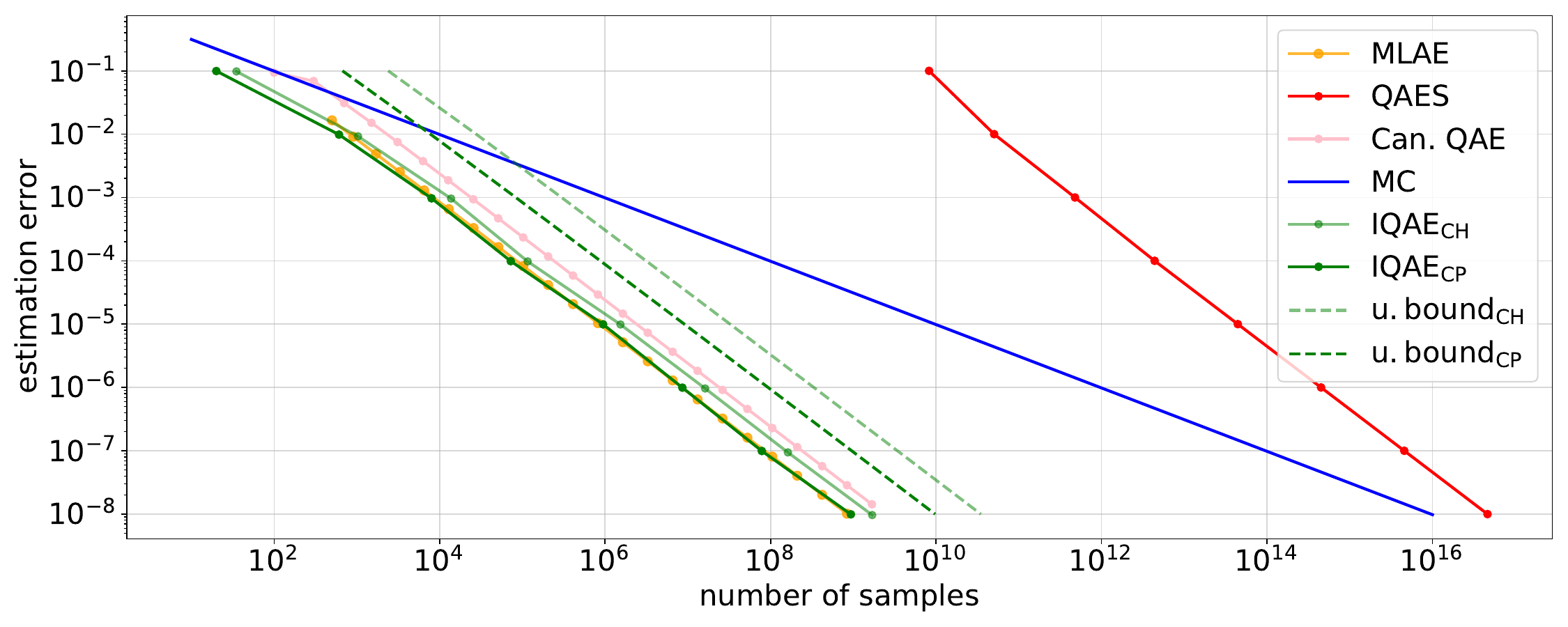}
\caption{Comparison of QAE variants: The resulting estimation error for $a=1/2$ and 95\% confidence level with respect to the required total number of oracle queries. We also include theoretical upper bounds for two versions of IQAE (CH = Chernoff-Hoeffding, CP = Clopper-Pearson). Note that QAES provides a relative error estimate, while the other algorithms return an absolute error estimate. For MLAE and canonical QAE we use the likelihood ratio confidence intervals. For MLAE we count the number of oracle calls for the largest power of $\mathcal{Q}$ operator, which corresponds to parallel execution of the algorithm. For IQAE, MLAE and canonical QAE we used $N_{\text{shots}}=100$. }
\label{fig:comparison_results}
\end{figure*}

To compare all algorithms we estimate $a = 1/2$ with a $1-\alpha = 95\%$ confidence interval.
For IQAE, MLAE, and QAE, we set $N_{\text{shots}} = 100$.
As shown in Fig.~\ref{fig:comparison_results}, IQAE outperforms all other algorithms.
QAES, even though achieving the best asymptotical behavior, performs worst in practice.
On average, QAES requires about $10^8$ times more oracle queries than IQAE which is even more than for classical MC simulation with the tested target accuracies. MLAE performs comparable to IQAE, however, the exact MLE becomes numerically challenging with increasing $m$. In order to observe the scaling of quantum part of the algorithm, we collect more data points via usage of geometrically smaller search domain around estimated $\theta$ with each new round instead of brute force search on the whole initial domain for $\theta$.
Lastly, QAE with MLE-postprocessing performs a bit worse than IQAE and MLAE, which answers the question raised at the beginning: Applying QPE in the QAE setting does not lead to any advantage but only increases the complexity, even with an MLE-postprocessing.
Thus, using IQAE instead does not only reduce the required number of qubits and gates, it also improves the performance.
Note that the MLE problem resulting from canonical QAE is significantly easier to solve than the problem arising in MLAE, since the solution can be efficiently computed with a bisection search, see Appendix \ref{sec:canonical_qae_with_mle}.
However, to evaluate QAE we need to simulate an increasing number of (ancilla) qubits, even for the very simple problem considered here, which makes the simulation of the quantum circuits more costly.

In the remainder of this section we analyze the performance of IQAE in more detail.
In particular, we empirically analyze the total number of oracle queries when using both Chernoff-Hoeffding the Clopper-Pearson confidence intervals as well as the resulting $k$-schedules.

\begin{figure}[hbtp]
\centering
\includegraphics[width=0.5\textwidth]{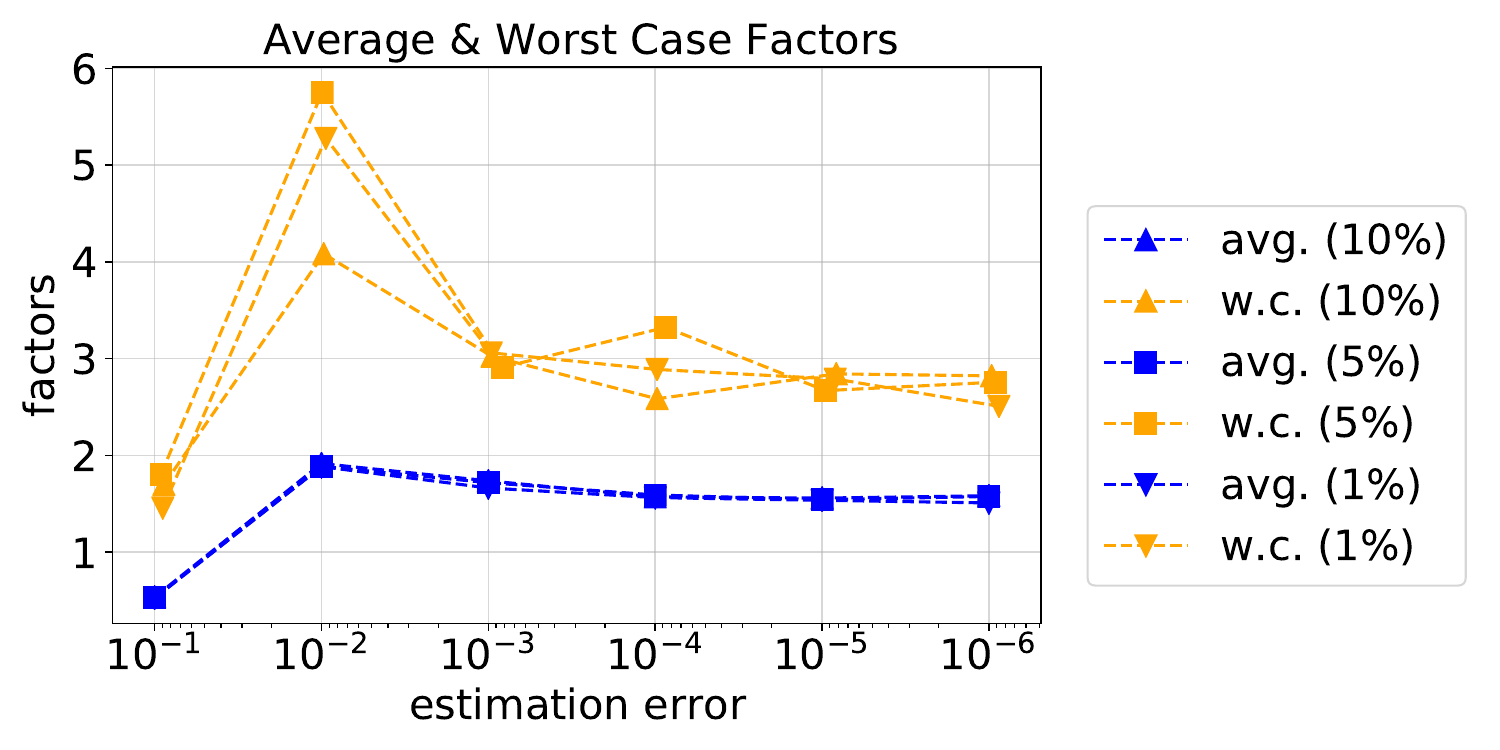}
\caption{Analysis of constant overhead for Chernoff-Hoeffding IQAE: The average (blue) and worst case (orange) constant overhead for IQAE runs with different parameter settings.}
\label{fig:iqae_ch_factors}
\end{figure}

\begin{figure}[hbtp]
\centering
\includegraphics[width=0.5\textwidth]{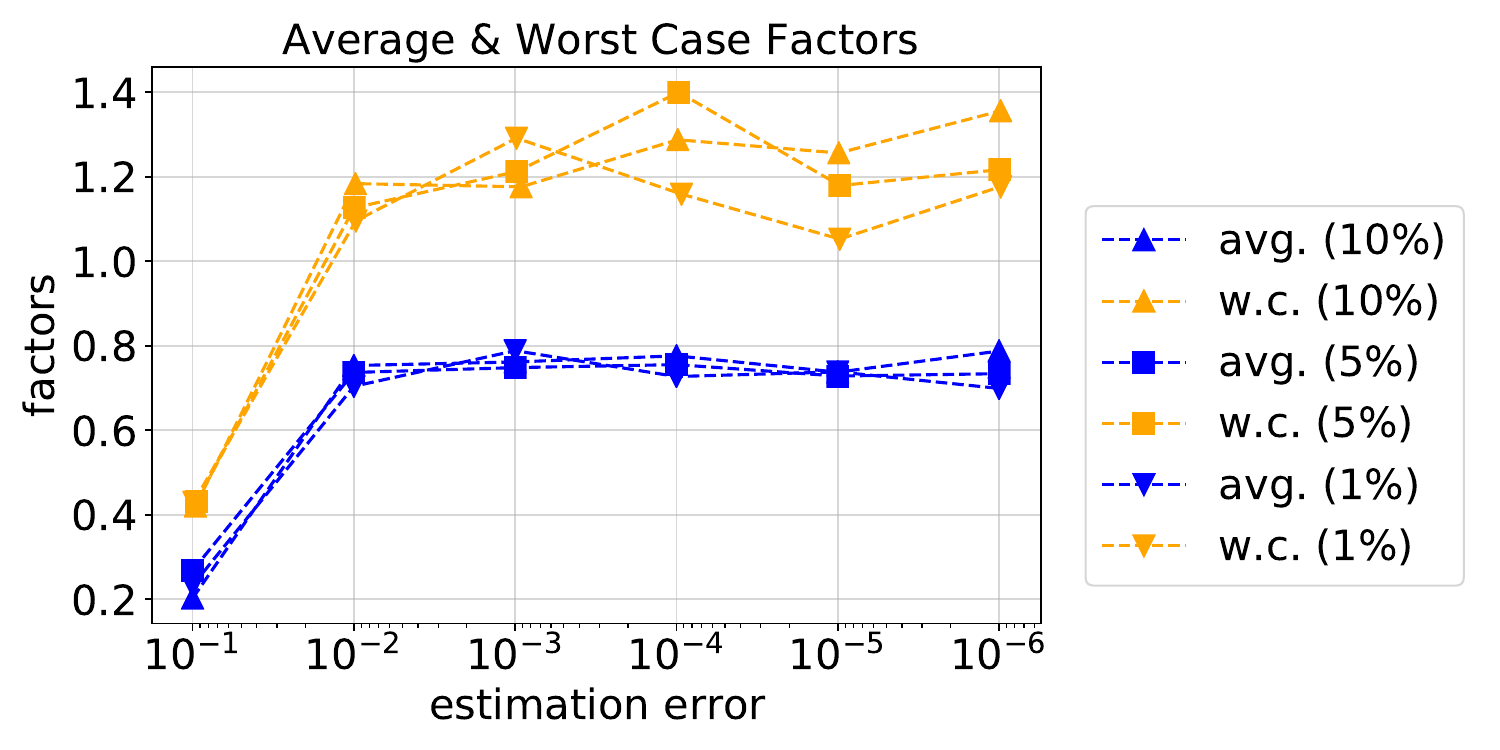}
\caption{Analysis of constant overhead for Clopper-Pearson IQAE.}
\label{fig:iqae_cp_factors}
\end{figure}

More precisely, we run IQAE for all $a \in \{i/100 \mid i = 0, \ldots, 100\}$ discretizing $[0, 1]$, for all $\epsilon \in \{10^{-i} \mid i = 3, \ldots, 6\}$, and for all $\alpha \in \{1\%, 5\%, 10\%\}$.
We choose $N_{\text{shots}} = 100$ for all experiments.
For each combination of parameters, we evaluate the resulting number of total oracle calls $N_{\text{oracle}}$ and compute

\begin{eqnarray}
\frac{N_{\text{oracle}}}{\log(2/\alpha \log_2(\pi/4\epsilon))/\epsilon},
\end{eqnarray}
 
i.e., the constant factor of the scaling with respect to $\epsilon$ and $\alpha$.
We evaluate the average as well as the worst case over all considered values for $a$.
The results are illustrated in Fig.~\ref{fig:iqae_ch_factors}-\ref{fig:iqae_cp_factors}. The empirical complexity analysis of Chernoff-Hoeffding IQAE leads to:

\begin{eqnarray}
N_{\text{oracle}}^{\text{avg}} &\leq& \frac{2}{\epsilon}\log\left(\frac{2}{\alpha} \log_2\left(\frac{\pi}{4\epsilon}\right)\right) \text{, and} \\
N_{\text{oracle}}^{\text{wc}}  &\leq& \frac{6}{\epsilon}\log\left(\frac{2}{\alpha} \log_2\left(\frac{\pi}{4\epsilon}\right)\right),
\end{eqnarray}

where $N_{\text{oracle}}^{\text{avg}}$ denotes the average and $N_{\text{oracle}}^{\text{wc}}$ the worst case complexity, respectively. Furthermore, the analysis of Clopper-Pearson IQAE leads to: 

\begin{eqnarray}
N_{\text{oracle}}^{\text{avg}} &\leq& \frac{0.8}{\epsilon}\log\left(\frac{2}{\alpha} \log_2\left(\frac{\pi}{4\epsilon}\right)\right) \text{, and} \\
N_{\text{oracle}}^{\text{wc}}  &\leq& \frac{1.4}{\epsilon}\log\left(\frac{2}{\alpha} \log_2\left(\frac{\pi}{4\epsilon}\right)\right),
\end{eqnarray}

\begin{figure}[hbtp]
\centering
\includegraphics[width=0.5\textwidth]{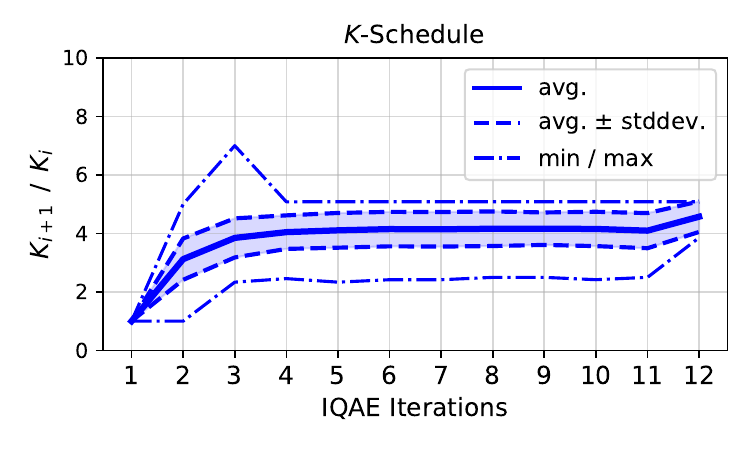}
\caption{$K$-schedule: Average, standard deviation, minimum, and maximum value of $K_{i+1} / K_{i}$ per iteration over $1,000$ repetitions of Clopper-Pearson IQAE for $a = 1/2$, $\epsilon = 10^{-6}$, $\alpha = 5\%$ and $N_{\text{shots}} = 100$.}
\label{fig:iqae_round_analysis}
\end{figure}

To analyze the $k$-schedule, we set $a = 1/2$, $\epsilon = 10^{-6}$, $\alpha = 5\%$ and again $N_{\text{shots}} = 100$.
Fig.~\ref{fig:iqae_round_analysis} shows for each iteration the resulting average, standard deviation, minimum, and maximum of $K_{i+1} / K_i$, over $1,000$ repetitions of the algorithm, for the $K_i$ defined in Alg.~\ref{alg:find_k}.
As explained in Section \ref{sec:iqae}, we want to achieve as high as possible value of $K_{i+1}/K_i$ for each iteration. Therefore, it can be seen that $N_{\text{shots}} = 100$ seems to be too small for the first round, i.e., another iteration with the same $K_i$ is necessary before approaching an average growth rate slightly larger than four. 

\section{Conclusion and Outlook} \label{sec:conclusion}

We introduced \emph{Iterative Quantum Amplitude Estimation}, a new variant of QAE that realizes a quadratic speedup over classical MC simulation.
Our algorithm does not require QPE, i.e., it is solely based on Grover iterations, and allows us to prove rigorous error and convergence bounds.
We demonstrate empirically that our algorithm outperforms the other existing variants of QAE, some even by several orders of magnitude.
This development is an important step towards applying QAE on quantum hardware to practically relevant problems and achieving a quantum advantage.

Our algorithm achieves the quadratic speedup up to a $\log(2/\alpha \log_2(\pi/4\epsilon))$-factor.
In contrast, QAES, the other known variant of QAE without QPE and with a rigorous convergence proof, achieves optimal asymptotic complexity at the cost of very large constants.
It is an open question for future research whether there exists a variant of QAE without QPE that is practically competitive while having an asymptotically optimal  performance bound.
Another difference between IQAE and QAES is the type of error bound: IQAE provides an absolute and QAES a relative bound.
Both types are relevant in practice, however, in the context of QAE, where problems often need to be normalized, a relative error bound is sometimes more appropriate.
We leave the question of a relative error bound for IQAE open to future research.

Another research direction that seems of interest is on the existence of parallel versions of QAE. More precisely, is it possible to realize the powers of the operator $\mathcal{Q}$ distributed somehow in parallel over additional qubits, instead of sequential application on the quantum register? However, as shown in  \cite{burchard2019lower}, this does not seem to be possible.

Another open question for further investigation is the optimal choice of parameters for IQAE.
We can set the required minimal growth rate for the oracle calls as well as the number of classical shots per iteration and both affect the performance of the algorithm.
Determining the most efficient setting may further reduce the required number of oracle calls for a particular target accuracy.

We also demonstrated in Sec.~\ref{sec:results}, that the gap between the bound on the total number of oracle calls provided in Thm.~\ref{thm:proof_of_convergence} and the actual performance is not too big. The proof technique for the upper bound almost achieves the actual performance. However, one may still ask whether an even tighter analytic bound is possible.

To summarize, we introduced and analyzed a new variant of QAE without QPE that outperforms the other known variants and we provide a rigorous convergence theory.
This helps to reduce the requirements on quantum hardware and is an important step towards leveraging quantum computing for real-world applications.

\section*{Data Availability}
The data that support the findings of this study are available from the corresponding author upon justified request.

\section*{Code Availability}

The mentioned algorithms are available open source as part of Qiskit and can be found in \url{https://github.com/Qiskit/qiskit/}. 
Tutorials explaining the algorithm and its application are located in \url{https://github.com/Qiskit/qiskit-tutorials}.

\section*{Acknowledgment}

D.G., J.G., and C.Z.~acknowledge the support of the National Centre of Competence in Research \textit{Quantum Science and Technology} (QSIT). D.G. also acknowledges the support from ESOP scholarship from 
ETH Zurich Foundation.

IBM, the IBM logo, and ibm.com are trademarks of International Business Machines Corp., registered in many jurisdictions worldwide. Other product and service names might be trademarks of IBM or other companies. The current list of IBM trademarks is available at \url{https://www.ibm.com/legal/copytrade}.

\section*{Author Contributions}

S.W.~conceived the idea and co-supervised the work together with C.Z. D.G.~developed IQAE and performed the theoretical analysis, J.G.~worked on the MLE post-processing for canonical QAE.
D.G.~and J.G.~jointly worked on the implementation and numerical experiments for all algorithms. D.G.~and S.W.~wrote the first draft of the manuscript and all authors contributed to its final version.

\section*{Competing Interests}
The authors declare that there are no competing interests.

\appendix

\section{\label{sec:canonical_qae_with_mle} Canonical QAE with MLE}

QAE can be enhanced by using a classical MLE postprocessing. 
Given a parametrized probability distribution $f$ with unknown parameter $\theta$ and data $\left\{x_i\right\}_i$, $i=1,\ldots, N$ sampled from it, MLE is a method to obtain an estimate $\hat{\theta}$ for $\theta$.
This is done by maximizing the likelihood $L$,
\begin{eqnarray}
    \hat\theta = \arg\max_{\theta^{\prime}} L(\theta^{\prime}) = \arg\max_{\theta^{\prime}} \prod_{i=1}^N f(x_i|\theta^{\prime}),
\end{eqnarray}
which is a measure for how likely it is to observe the data $\left\{x_i\right\}_i$, given that  $\theta^{\prime}$ is the true parameter.
Numerically it is often favourable to maximize the log-likelihood $\log{L}$.

\begin{figure}[hbtp]
\centering
\includegraphics[width=0.5\textwidth]{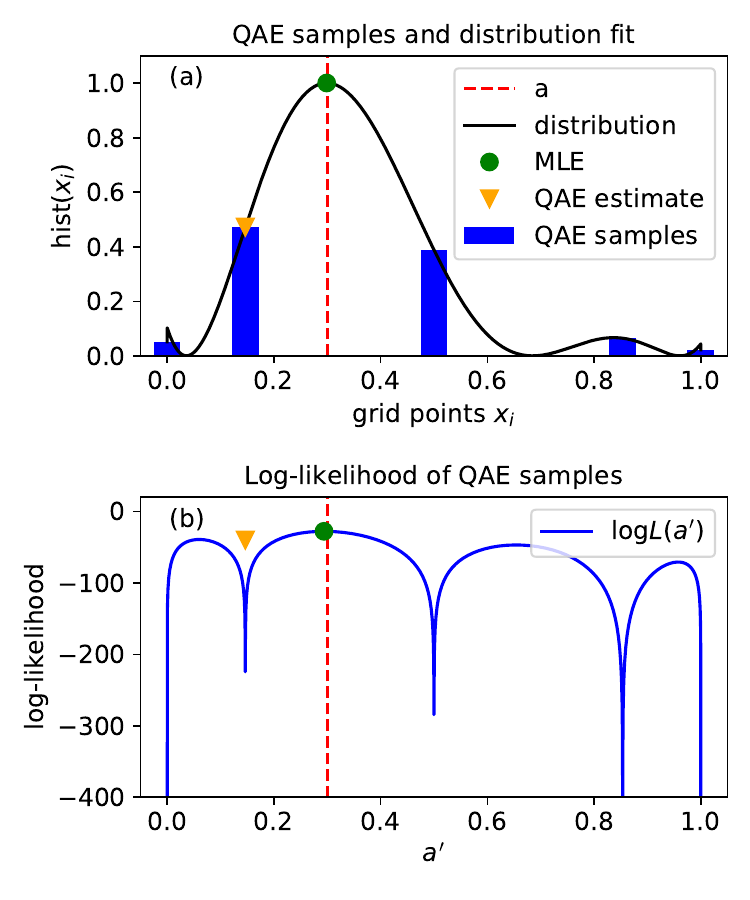}
\caption{MLE postprocessing for QAE: In this problem, we set $a=0.3$, $m=3$ and used 25 shots for QAE.
(a) The distribution (black line) is fitted to the the normalized histogram of the  QAE samples (blue bars). The QAE output is the median of all samples (orange triangle). The MLE (green dot) in the QAE setting is also the peak of the distribution and very close to the real $a$. 
(b) The MLE is the global maximum of the log-likelihood function (blue line). The search for the MLE is conducted as a bisection search in the two neighbouring bubble-shaped intervals of the QAE estimate.}  
\label{fig:qae_mle}
\end{figure}

In QAE, we try to approximate $a$ with the MLE $\hat{a}$.
The probability distribution $f$ is the probability to sample a certain grid point in one measurement. From \cite{brassard}, this distribution can be derived to be
\begin{equation}
f(x_i | a^{\prime}) =
\begin{cases} 
        \mathbb{P}[\ket{y(a^{\prime})}] + \mathbb{P}[\ket{M - y(a^{\prime})}], &\text{if } a^{\prime} \not\in \{0, 1\} \\
        \mathbb{P}[\ket{y(a^{\prime})}], &\text{otherwise }
        \end{cases},
\end{equation}
with $y(a^{\prime}) = M\arcsin{}(\sqrt{a^{\prime}}) / \pi$,  $\mathbb{P}[\ket{y}]$ from Eq.~\ref{eq:qae_sample_probability}, and grid points $x_i = \sin^2\left(i\pi/M\right), i \in \{0, \ldots, M/2 \}$. 
See Fig.~\ref{fig:qae_mle}a for a visualization of the probability distribution fitted to the QAE samples and Fig.~\ref{fig:qae_mle}b for the respective log-likelihood function $\log L$.

To find the maximum of $\log L$ without much overhead, we exploit the information given by the QAE output: if QAE is successful its estimate is the closest grid point to $a$. 
Thus, we only have to search the intervals of the neighbouring grid points to find the exact $a$, which is done using a bisection search.
Note that for $N \rightarrow \infty$, this search would return the exact amplitude, i.e. $\hat a = a$, independent of the number of qubits $m$.

Confidence intervals for the MLE can be derived using the Fisher information in combination with the central-limit theorem or with the likelihood ratio (LR) \cite{koch_1999_parameter}.
In our tests, the LR was more reliable than other approaches such as (observed) Fisher information. Due to the data-based definition of the LR confidence intervals, the better performance fits the expectations \cite{Maldonado1994, Jeng2000}.
The LR confidence interval uses the fact that for large sample numbers $N$ the LR statistic is approximately $\chi^2$-distributed, with one degree of freedom:
\begin{eqnarray}
    2\log\frac{L(\hat a)}{L(a)} \sim \chi^2_1.
\end{eqnarray}
The statistic can be used to conduct a two-sided hypothesis test with the null hypothesis $H_0: \hat a = a$, which is rejected at the $\alpha$ level if the LR statistic exceeds the $(1 - \alpha)$ quantile of the $\chi^2$ distribution, $q_{\chi_1^2}(1 - \alpha)$. 
The corresponding confidence interval is $\left\{ a^{\prime} \in [0, 1] : \log L(a^{\prime}) \geq \log L(\hat{a}) - q_{\chi^2_1}(1 - \alpha) / 2\right\}$.
Using the likelihood function for the QAE samples we obtain confidence intervals for the MLE $\hat a$ of $a$.
We apply the same approach to derive confidence intervals for MLAE.

\section{\label{sec:proof_of_convergence} Proof of Theorem \ref{thm:proof_of_convergence}}
Let us outline the strategy. We are going to use the union bound to combine the estimates, which are derived from the Chernoff-Hoeffding bound, for different rounds of the algorithm and an upper bound $T$ for the actual number of rounds $t$ required to derive an upper bound for the total query complexity in terms of the desired precision $\epsilon$ for the parameter $a$ and confidence level $1-\alpha$.

\begin{proof}
Suppose a given confidence interval $[a_l, a_u]$ for $a$, and recall that $a = \sin^2(\theta_a)$.
This implies
\begin{eqnarray}
\frac{\abs{a_u-a_l}}{2}
&=& \frac{\abs{\sin(\theta_u+\theta_l)} \abs{\sin(\theta_u-\theta_l)}}{2} \label{eq:athetaeps} \\
&\leq& \frac{\abs{\theta_u-\theta_l}}{2}. \nonumber
\end{eqnarray}
Thus, to achieve $\abs{a_u - a_l}/2 \leq \epsilon$, it suffices to achieve $\abs{\theta_u-\theta_l}/2 \leq \epsilon$ for our estimate of $\theta_a$.

Suppose that in round $i$ our knowledge of $\theta_a$ is the confidence interval $[\theta_l,\theta_u]$, and we just applied the Grover operator $k_i$ times.
Recall that the application of $\mathcal{Q}^{k_i}$ to $\mathcal{A}\ket{0}$ effectively multiplies the angle $\theta_a$ by $K_i = 4 k_i + 2$.
Denote $\theta_i^{\text{min}} \defeq \{K_i \theta_l \}_{\text{mod }2\pi}$ and $\theta_i^{\text{max}} \defeq \{K_i \theta_u\}_{\text{mod }2\pi}$, where, unlike in Alg.~\ref{alg:find_k}, we use fractional part modulo $2\pi$ of a given number. These are determined by measuring the variable $a_i$: their estimates $\tilde{a}_i$ for each round $i$ define the approximated probabilities of positive outcomes that we see from measurements and the corresponding confidence intervals $[a_i^{\text{min}}, a_i^{\text{max}}]$:
\begin{equation}
\begin{aligned}
&a_i \defeq (1 - \cos(K_i\theta_a))/2, \\
&a_i^{\text{min}} \defeq \max(0, \tilde{a}_i - \tilde{\epsilon}_{a_i}), \\
&a_i^{\text{max}} \defeq \min(1, \tilde{a}_i + \tilde{\epsilon}_{a_i}), \\
&\epsilon_{a_i} \defeq \abs{a_i^{\text{max}} - a_i^{\text{min}}}/2,
\label{eq:aidefs}
\end{aligned}
\end{equation}
where $\tilde{\epsilon}_{a_i}$ is a half-width of a confidence interval, calculated from Chernoff-Hoeffding bound for i.i.d.~Bernoulli trials with $N_{\text{shots}}$ samples \cite{hoeffding1963probability}, which is given by
\begin{equation}
\mathbb{P}[a_i \notin [a_i^{\text{min}},a_i^{\text{max}}]] \leq 2 \exp \left(-2 N_{\text{shots}}\tilde{\epsilon}_{a_i}^2\right).
\label{eq:chernoff}
\end{equation}

Note, that actual error $\epsilon_{a_i}$ can be lower than $\tilde{\epsilon}_{a_i}$. If this scenario is realized, then $\theta_i^{\text{max}} \in \{\pi,2\pi\}$ or $\theta_i^{\text{min}} \in \{0,\pi\}$, i.e. the confidence interval touches the boundary of the upper or lower half-circle.

Since we are looking for a sufficient number of measurement shots to ensure the algorithm runs correctly, we are going to refer to the number of shots as $N_{\text{max}}$ from now on. If we require
\begin{equation}
2 \exp(-2N_{\text{max}} \tilde{\epsilon}_{a_i}^2) = \frac{\alpha}{T} , \quad \forall i \in \{1,\ldots,t\},
\label{eq:chernoffcond}
\end{equation}
then the union bound asserts that the main condition of the theorem -- the guarantee for the total error probability being bounded by $\alpha$ -- is satisfied:
\begin{equation}
\begin{aligned}
& \mathbb{P}[a \notin [a_l, a_u]] \\
& \leq \mathbb{P}\left[ \exists i \in \{1,\ldots,t\}: a_i \notin [a_i^{\text{min}},a_i^{\text{max}}] \right] \\ 
& \leq \sum_{i=1}^{t}  \mathbb{P}[a_i \notin [a_i^{\text{min}},a_i^{\text{max}}]] \\
& \leq \sum_{i=1}^{t} 2 \exp(-2N_{\text{max}} \tilde{\epsilon}_{a_i}^2) =  t\frac{\alpha}{T} \leq \alpha.
\end{aligned}
\end{equation}
From (\ref{eq:chernoffcond}) and $\epsilon_{a_i} \leq \tilde{\epsilon}_{a_i}$ we deduce:
\begin{eqnarray}
\epsilon_{a_i}^2 \leq \tilde{\epsilon}_{a_i}^2 = \frac{1}{2N_{\text{max}}}\log(\frac{2T}{\alpha}) , \quad \forall i \in \{1,\ldots,t\}
\label{eq:epsN}
\end{eqnarray}

Now let us derive an upper bound $T$ for the actually required number of rounds $t$ to achieve the desired absolute error $\epsilon$ for $\theta_a$. First, given the actual schedule $\{q_i\}_{i=1}^{t-1}$, we have the following trivial relation for the last step of the algorithm:
\begin{eqnarray}
\frac{L_{\text{min}}}{K_t} \leq \epsilon < \frac{L_{\text{max}}}{K_{t-1}}.
\label{eq:prodqidef}
\end{eqnarray}
where $L_{\text{min}}$ and $L_{\text{max}}$ are lower and upper bounds for $\epsilon_{\theta_t}$ respectively (see (\ref{def:LminCH})-(\ref{def:LmaxCH})) and $K_t = K_1 \prod_{i=1}^{t-1}q_i$. Moreover, the "no-overshooting" condition of Alg.~\ref{alg:iqae} gives us a stronger bound for $N_t K_t$:
\begin{equation}
N_t K_t \leq N_{\text{max}} \frac{L_{\text{max}}}{\epsilon},
\label{eq:Kt}
\end{equation} 
where $N_t$ stands for the number of shots at the last round of the algorithm.
Secondly, we define an integer $T = T(r)$ for $r > 1$ as follows:
\begin{eqnarray}
 \frac{L_{\text{max}}}{K_1 r^{T-1}} \leq \epsilon < \frac{L_{\text{max}}}{K_1 r^{T-2}}.
\label{eq:Tdef}
\end{eqnarray}
i.e. $T \defeq \ceil{\log_r(rL_{\text{max}}/2\epsilon) }$ (we used $K_1 = 2$). Note, that this definition together with (\ref{eq:prodqidef}) and (\ref{eq:qi})  leads to the following estimate:
\begin{eqnarray}
r^{T-1-j} > \prod_{i=1}^{t-2-j}q_i, \quad \forall j \in \{0,\ldots,t-2\},
\label{eq:qvsT}
\end{eqnarray}
which itself implies $r^{T-1} > r^{t-2}$, i.e $T \geq t$ and $T$ is an upper bound for $t$ as it should be.

If we define parameter $L \in [0,\pi/2]$ as
\begin{eqnarray}
\sin^2(L) &\defeq & 2\tilde{\epsilon}_{a_i},
\label{eq:Ldef}
\end{eqnarray}
it follows from (\ref{eq:epsN}) that maximum number of shots required is given by
\begin{eqnarray}
N_{\text{max}}(\epsilon, \alpha) &=& \frac{2}{\sin^4(L)} \log(\frac{2T(\epsilon)}{\alpha}) \nonumber \\ 
&\leq& \frac{2}{\sin^4(L)} \log(\frac{2}{\alpha} \log_r \left (\frac{r^2L_{\text{max}}}{2\epsilon} \right)). \nonumber \\
\label{eq:n_max_bound}
\end{eqnarray}

Lemma~\ref{thm:lemma}, which can be found below, finds the optimal value of $L$, denoted further in the proof by $L^*$, such that on every round $i$ we have the condition $\forall r \in (1,3] \text{ } q_{i} \geq r$.

Finally, we derive a bound for the total number of $\mathcal{Q}$-oracle calls, denoted by $N_{\text{oracle}}$, using (\ref{eq:Kt}), (\ref{eq:qvsT}) and (\ref{eq:qi}):
\begin{eqnarray}
&&N_{\text{oracle}} = \sum_{i=1}^{t} N_i k_i = \sum_{i=1}^{t} N_i \frac{K_i - 2}{4} \nonumber \\
&&\leq \frac{N_{\text{max}}}{2} \left(1 - t + \frac{K_1}{2} \sum_{i=1}^{t-2}  \prod_{j=1}^{i}q_j \right) + \frac{N_t K_{t}}{2} \nonumber \\
&&< \frac{N_{\text{max}}}{2} \left(1 - t + \sum_{i=1}^{t-2} r^{T-t+1+i} + \frac{L_{\text{max}}}{2\epsilon} \right) \nonumber \\
&&= \frac{N_{\text{max}}}{2} \left( 1 - t + r^{T-t+1} \frac{r^{t-1} - r}{r-1} +\frac{L_{\text{max}}}{2\epsilon}\right)  \nonumber \\
&&= \frac{N_{\text{max}}}{2} \left(r^{T-2}\frac{r^2}{r-1} + \frac{L_{\text{max}}}{2\epsilon}  +  1 - t - \frac{r^{T-t+2}}{r-1} \right)  \nonumber \\
&&< \frac{N_{\text{max}}}{2} \left( \frac{L_{\text{max}}}{2\epsilon} \frac{r^2}{r-1} + \frac{L_{\text{max}}}{2\epsilon} + 3 - t - \frac{r^{T-t+2}}{r-1} \right)  \nonumber \\
&&< N_{\text{max}} \frac{L_{\text{max}}}{4\epsilon} \left( 1 +  \frac{r^2}{r-1} \right).
 \label{eq:n_oracle_bound}
\end{eqnarray}
Combining (\ref{eq:n_max_bound}) with Lemma~\ref{thm:lemma} and (\ref{eq:n_oracle_bound}), we get
\begin{eqnarray}
N_{\text{oracle}} &<& \frac{1}{\epsilon} \times  \frac{L_{\text{max}}}{2\sin^4(L_{\text{max}})}  \left( 1 +  \frac{r^2}{r-1} \right) \nonumber \\ 
&& \times \log(\frac{2}{\alpha} \log_r \left (\frac{r^2 L_{\text{max}}}{2\epsilon} \right)),  \nonumber
\end{eqnarray}
which, using $r=2$, $L_{\text{max}} = L^* < 11\pi/90 < \pi/8$ and $\sin^4(L_{\text{max}}) = (1/4-\sin(\pi/14)/2)^2$, gives us the bound
\begin{eqnarray}
N_{\text{oracle}} &<& \frac{1}{\epsilon} \times \frac{44\pi}{9(1-2\sin(\pi/14))^2} \times \log(\frac{2}{\alpha} \log_2 \left (\frac{\pi}{4\epsilon} \right)). \nonumber
\end{eqnarray}
This inequality implies the following numerical form:
\begin{eqnarray}
N_{\text{oracle}} 
&<& \frac{50}{\epsilon} \log(\frac{2}{\alpha}\log_2 \left (\frac{\pi}{4\epsilon}  \right)). \nonumber
\end{eqnarray}
\end{proof}

\begin{figure*}[hbtp]
\centering
\includegraphics[width=\textwidth]{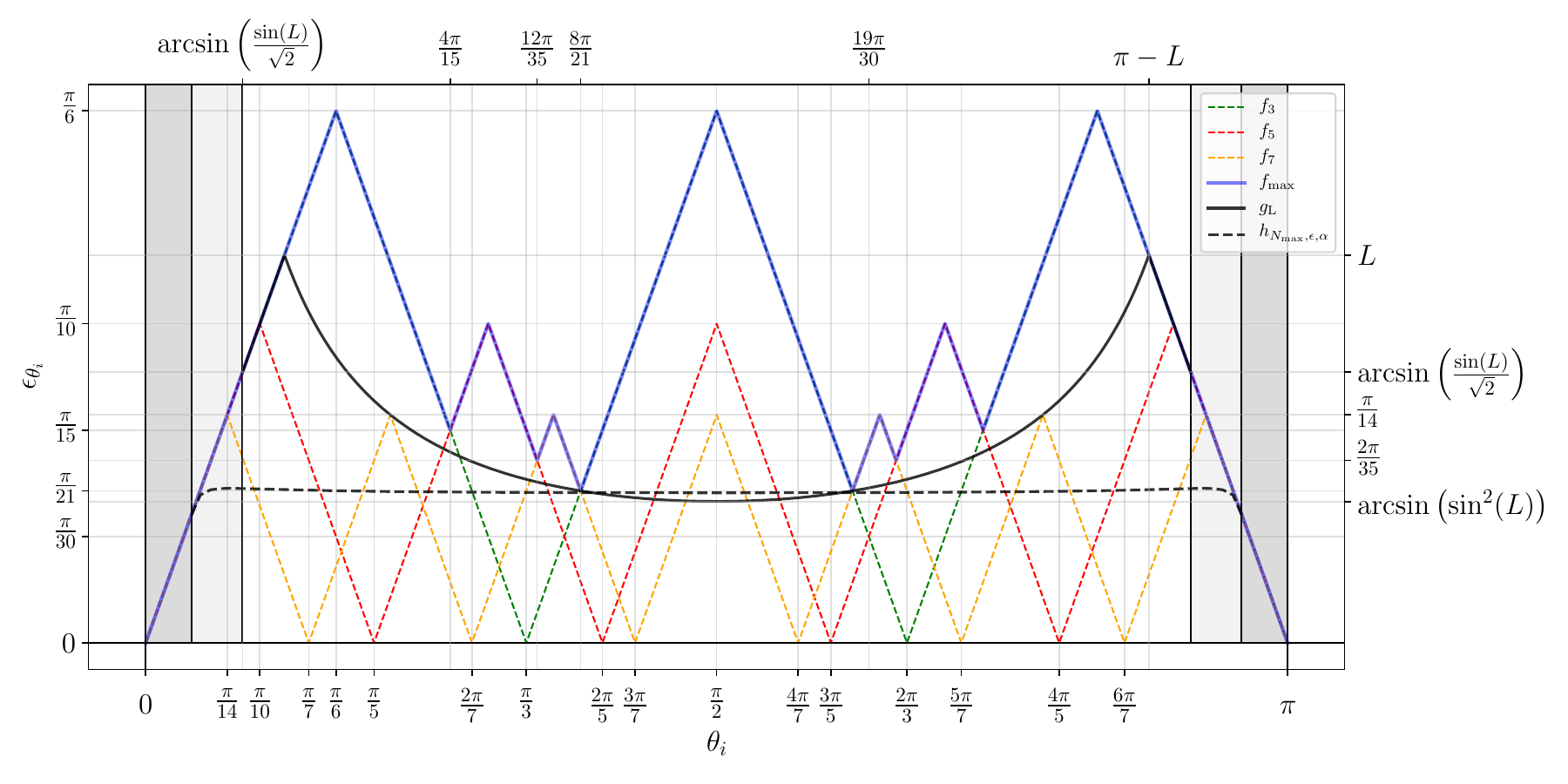}
\caption{Functions $f_3, f_5, f_7$ define the regions of validity for corresponding conditions (\ref{eq:sec3}),(\ref{eq:sec5}) and (\ref{eq:sec7}) respectively. The area under function $f_{\text{max}}$ represents the logical union of these conditions. Finally, if we choose to use $N_{\text{max}}(L)$ shots for the measurement, the function $g_L(\theta_i)$ represents the behavior of confidence interval width, which apparently depends on the estimated value $\theta_i$. Shaded regions are not included in the domain of $g_L(\theta_i)$, because the estimates $\theta_i$ cannot lie inside them for a chosen number of shots. The same is true for the case of Clopper-Pearson bound (see Appendix~\ref{sec:proof_of_convergence_CP}), which corresponds to the function $h_{N_{\text{max}},\epsilon,\alpha}$.}
\label{fig:lemma}
\end{figure*}

\begin{lemma}[Sufficient condition for $N_{\text{max}}$]\label{thm:lemma}
For any $L~\leq~L^*$, where $L^*= \arcsin(\frac{1}{2}\sqrt{1-2\sin(\frac{\pi}{14})})$ the number of shots $N_{\text{max}}(L)$, defined by (\ref{eq:n_max_bound}), is sufficient to ensure 
\begin{eqnarray}
\forall r \in (1,3] \text{ } q_{i} \geq r.
\label{eq:qi}
\end{eqnarray}
Moreover, $L_{\text{max}} = L$ and $L_{\text{min}} = \arcsin(\sin^2(L))$.
\begin{proof} 
Given a confidence interval $[\theta_i^{\text{min}},\theta_i^{\text{max}}]$, it is equivalent to look at it as a pair of midpoint and error $\{\theta_i, \epsilon_{\theta_i}\}$, but we need
to translate $\epsilon_{a_i}$ into $\epsilon_{\theta_i}$ first. Analogously to (\ref{eq:athetaeps}) we can write
\begin{eqnarray}
\epsilon_{a_i} &=& \frac{\abs{\sin((\theta_i^{\text{max}}+\theta_i^{\text{min}})/2)} \abs{\sin((\theta_i^{\text{max}}-\theta_i^{\text{min}})/2)}}{2} \nonumber \\
 &=& \frac{\abs{\sin(\theta_i)}\abs{\sin(\epsilon_{\theta_i})}}{2},
\label{eq:epsai}
\end{eqnarray}
where $\theta_i \defeq (\theta_i^{\text{max}}+\theta_i^{\text{min}})/2$. 
Now using (\ref{eq:epsai}) and (\ref{eq:Ldef}) we can understand how to calculate the error $\epsilon_{\theta_i}$ from $\epsilon_{a_i}$ (or equivalently from $L$). For that purpose we introduce a family of functions $g_{L}(\theta_i)$
\begin{equation}
\begin{aligned}
&g_L(\theta_i) \defeq\\
&\begin{cases}
        \min \left( \arcsin \left( \frac{\sin^2(L) }{\sin(\theta_i)} \right),  \theta_i, \pi - \theta_i \right), & \theta_i \in \text{dom}^{u}(g_L) \\
       \min \left( \arcsin \left( \frac{\sin^2(L) }{\sin(\theta_i - \pi)} \right), \theta_i - \pi, 2\pi - \theta_i \right), & \theta_i \in \text{dom}^{d}(g_L)
        \end{cases}
\label{eq:errorCalc}
\end{aligned}
\end{equation}
and we have $\epsilon_{\theta_i} = g_{L}(\theta_i)$, where the domain of $g_L$ depends on $L$ and equals $\text{dom}(g_L) =  \text{dom}^{u}(g_L) \cup \text{dom}^{d}(g_L)  $ with 
$$\text{dom}^{u}(g_L)=\left[\arcsin(\frac{\sin(L)}{\sqrt{2}}),\pi - \arcsin(\frac{\sin(L)}{\sqrt{2}})\right] $$
and 
$$\text{dom}^{d}(g_L)=\left[\pi + \arcsin(\frac{\sin(L)}{\sqrt{2}}),2\pi - \arcsin(\frac{\sin(L)}{\sqrt{2}})\right]$$

Note, that the structure of domains reflects in itself the corner cases, when $\theta_i^{\text{max}} \in \{\pi,2\pi\}$ or $\theta_i^{\text{min}} \in \{0,\pi\}$. 
Moreover, the function $g_L$ encodes nicely boundary cases when $\epsilon_{a_i} < \tilde{\epsilon}_{a_i}$  (i.e. when $2\epsilon_{a_i} < \sin^2(L)$).
From the behavior of $g_{L}$ we also see that smallest possible error $\epsilon_{\theta_i}$ is attained at $\pi/2$ and equals 
\begin{equation}
L_{\text{min}} = \arcsin(\sin^2(L)). 
\label{def:LminCH}
\end{equation}
Analogously, 
\begin{equation}
L_{\text{max}} = L
\label{def:LmaxCH}
\end{equation}
and $\epsilon_{\theta_i} =L_{\text{max}}$ when $\theta_i = L$.

Observe now, that if there exists $j \in \{0, \ldots, 5\}$, such that 
\begin{eqnarray}
[\theta_i^{\text{min}}, \theta_i^{\text{max}}] \subseteq \left[ \frac{j\pi}{3}, \frac{(j+1)\pi}{3} \right], 
\label{eq:sec3}
\end{eqnarray}
then we can always find a new $K_{i+1}$ with $q_{i} \defeq K_{i+1}/K_i \geq 3$.
Analogously, if there exists $j \in \{0, \ldots, 9\}$, such that
\begin{eqnarray}
[\theta_i^{\text{min}}, \theta_i^{\text{max}}] \subseteq \left[ \frac{j\pi}{5}, \frac{(j+1)\pi}{5} \right],
\label{eq:sec5}
\end{eqnarray}
then we can always find a new $K_{i+1}$, such that $q_{i} \geq 5$. Finally, if there exists $j \in \{0, \ldots, 13\}$, such that
\begin{eqnarray}
[\theta_i^{\text{min}}, \theta_i^{\text{max}}] \subseteq \left[ \frac{j\pi}{7}, \frac{(j+1)\pi}{7} \right],
\label{eq:sec7}
\end{eqnarray}
then we can always find a new $K_{i+1}$, such that $q_{i} \geq 7$.
In all of the above cases, we have $q_{i} \geq 3$. In particular,  the condition (\ref{eq:qi}) is satisfied. 

In order to study when this happens we define piecewise linear functions $f_i$ on $[0,\pi]$ for every integer $i>1$:
\begin{eqnarray}
&f_i(x) \defeq  \min\left(\frac{(j-1)\pi}{i}+x,\frac{j\pi}{i} - x\right) \nonumber \\ &\text{for every $j \in \{1,...,i\}$ on every interval } \left[\frac{(j-1)\pi}{i}, \frac{j\pi}{i}\right]  \nonumber \\
\label{eq:f_i}
\end{eqnarray}
On $[\pi,2\pi]$ functions $f_i$ are defined analougosly by symmetry. These functions represent the distance from a given point on circle to the closest point from a set $\{\frac{j\pi}{i}\}_{j=0}^{2i}$. So we can consider functions $f_3, f_5, f_7$, which correspond to conditions (\ref{eq:sec3}),(\ref{eq:sec5}) and (\ref{eq:sec7}) respectively. To ensure that at least one of this conditions is satisfied in a given round $i$, we can require that
\begin{eqnarray}
g_L(\theta_i) \leq f_{\text{max}}(\theta_i) \text{ for every $\theta_i \in \text{dom}(g_L)$, }
\label{eq:fmax_cond}
\end{eqnarray}
where $f_{\text{max}}(\theta_i) \defeq \max(f_3(\theta_i), f_5(\theta_i), f_7(\theta_i))$.
The behavior of all functions is given in Fig.~\ref{fig:lemma} for the top half of the circle - the bottom half is symmetric. Trivial analysis of these functions leads to a condition equivalent to (\ref{eq:fmax_cond}):
\begin{eqnarray}
g_{L}\left(\frac{8\pi}{21}\right) \leq f_{\text{max}}\left(\frac{8\pi}{21}\right) =\frac{\pi}{21}
\label{eq:fmax_cond_equiv}
\end{eqnarray}
and corresponding equation for $L^*$:
\begin{eqnarray}
g_{L^*}\left(\frac{8\pi}{21}\right) = \frac{\pi}{21}
\label{eq:fmax_cond_L*}
\end{eqnarray}
which gives us the formula
\begin{eqnarray}
L^*= \arcsin(\frac{1}{2}\sqrt{1-2\sin(\frac{\pi}{14})}).
\label{eq:L*}
\end{eqnarray}
\end{proof}
\end{lemma}

\remark{We emphasize, that the proof provides an upper bound on the number of shots $N_{\text{max}}$, which are sufficient to run the algorithm. This was essentially done to ensure that condition $q_i \geq r$ is satisfied, which itself allowed us to prove the upper bound. In practice, we can make less number of measurements $N_{\text{shots}} < N_{\text{max}}$ in a given \textit{iteration} until Alg.~\ref{alg:find_k} finds $q_i$ big enough to progress onto the next round. Because of that, a given \textit{round} may consist of several iterations of the $while$-loop in Alg.~\ref{alg:iqae}. Furthermore, we could get better constants in the proof if we use Clopper-Pearson interval bound instead of the Chernoff-Hoeffding bound. However, analytic treatment of this case is either impossible or highly involved, so we instead provide numerical evidence for the better performance of the Clopper-Pearson case in Sec.~\ref{sec:results}, and a correctness theorem \ref{thm:proof_of_convergence_CP} for a fixed range of precision parameter $\epsilon$ and level $\alpha$ via numerical method in Appendix~\ref{sec:proof_of_convergence_CP}.}

\section{\label{sec:proof_of_convergence_CP} A variant of the theorem \ref{thm:proof_of_convergence} for Clopper-Pearson interval method}

Here we want to use Clopper-Pearson confidence interval \cite{ClopperPearson1934,scholz2008confidence} instead of Chernoff-Hoeffding bound. Since it is impossible to obtain an analytic proof for this case, we will use a numerical calculations in the proof and fixed confidence level of $0.95$, which is the most common one in practice.

\begin{theorem}[Correctness of IQAE with usage of Clopper-Pearson confidence interval]\label{thm:proof_of_convergence_CP}
Suppose a confidence level $1 - \alpha = 0.95$, a target accuracy $\epsilon \geq 2^{-200}$, and a number of shots $N_{\text{shots}} \in \{1, ..., N_{\text{max}}(\epsilon, \alpha)\}$, where 
$$N_{\text{max}}(\epsilon, \alpha) = 69 \log(\frac{2}{\alpha} \log_2 \left( \frac{\pi}{4\epsilon} \right)).$$

In this case, IQAE (Alg.~\ref{alg:iqae}) terminates after a maximum number of $\ceil{\log_2(\pi/8\epsilon)}$ rounds, where we define one round as a set of iterations with the same $k_i$, and each round consists of at most $N_{\text{max}}(\epsilon, \alpha) / N_{\text{shots}}$ iterations.
IQAE computes $[\theta_l, \theta_u]$ with $\theta_u - \theta_l \leq 2\epsilon$ and
\begin{eqnarray}
\mathbb{P}[\theta_a \notin [\theta_l, \theta_u]] \leq \alpha,
\end{eqnarray}
and returns $[a_l, a_u]$ with $a_u - a_l \leq 2\epsilon$ and
\begin{eqnarray}
\mathbb{P}[a \notin [a_l, a_u]] \leq \alpha.
\end{eqnarray}
Thus, $\tilde{a} = (a_l + a_u)/2$ leads to an estimate for $a$ with $|a - \tilde{a}| \leq \epsilon$ with a confidence of $1-\alpha$.

Furthermore, for the total number of $\mathcal{Q}$-applications, $N_{\text{oracle}}$, it holds that
\begin{eqnarray}
N_{\text{oracle}} 
&<& \frac{14}{\epsilon} \log(\frac{2}{\alpha}\log_2 \left (\frac{\pi}{4\epsilon}  \right))
\end{eqnarray}
\end{theorem}

\begin{proof}
The proof is completely analogous to Theorem~\ref{thm:proof_of_convergence}, except that instead of Chernoff-Hoeffding $1-\alpha$ confidence interval we have different Clopper-Pearson condition for $[a^{\text{min}},a^{\text{max}}]$:
\begin{equation}
\sup_{a} \mathbb{P}[a \notin [a^{\text{min}}(X,n,\alpha),a^{\text{max}}(X,n,\alpha)]] = \alpha
\label{eq:clopper-pearson}
\end{equation}
where $X$ is Binomial random variable, constructed from $n$ Bernoulli trials with probability parameter $a$. This condition is satisfied for the following functions \cite{scholz2008confidence}:
\begin{eqnarray}
&a^{\text{min}}(x,n,\alpha) \defeq I^{-1}(1 - \alpha/2 ; x + 1 , n - x) \label{eq:cp_min_def}\\
&a^{\text{max}}(x,n,\alpha) \defeq I^{-1}(\alpha/2 ; x , n - x + 1)
\end{eqnarray}
where $I^{-1}(\alpha; x, y)$ is the inverse of regularized incomplete beta function, i.e. it is a solution $p$ of the equation $I_p(x , y) = \alpha$, where $I_p(x , y)$ is regularized incomplete beta function.

For each round  Clopper-Pearson condition reads as
\begin{equation}
\sup_{a_i} \mathbb{P}\left[a_i \notin \left[a_i^{\text{min}},a_i^{\text{max}}\right]\right] = \frac{\alpha}{T(\epsilon)}
\label{eq:clopper-pearson-i}
\end{equation}
where
\begin{eqnarray}
a_i^{\text{min}} \defeq a^{\text{min}}\left(X_i,N_{\text{max}},\frac{\alpha}{T(\epsilon)}\right) \nonumber \\
a_i^{\text{max}} \defeq a^{\text{max}}\left(X_i,N_{\text{max}},\frac{\alpha}{T(\epsilon)}\right) \nonumber
\end{eqnarray}
and $X_i$ are Binomial random variables for each round, representing the measured outcomes and $T(\epsilon)$ is given by (\ref{eq:Tdef}).

\begin{figure*}[hbtp]
\centering
\includegraphics[width=\textwidth]{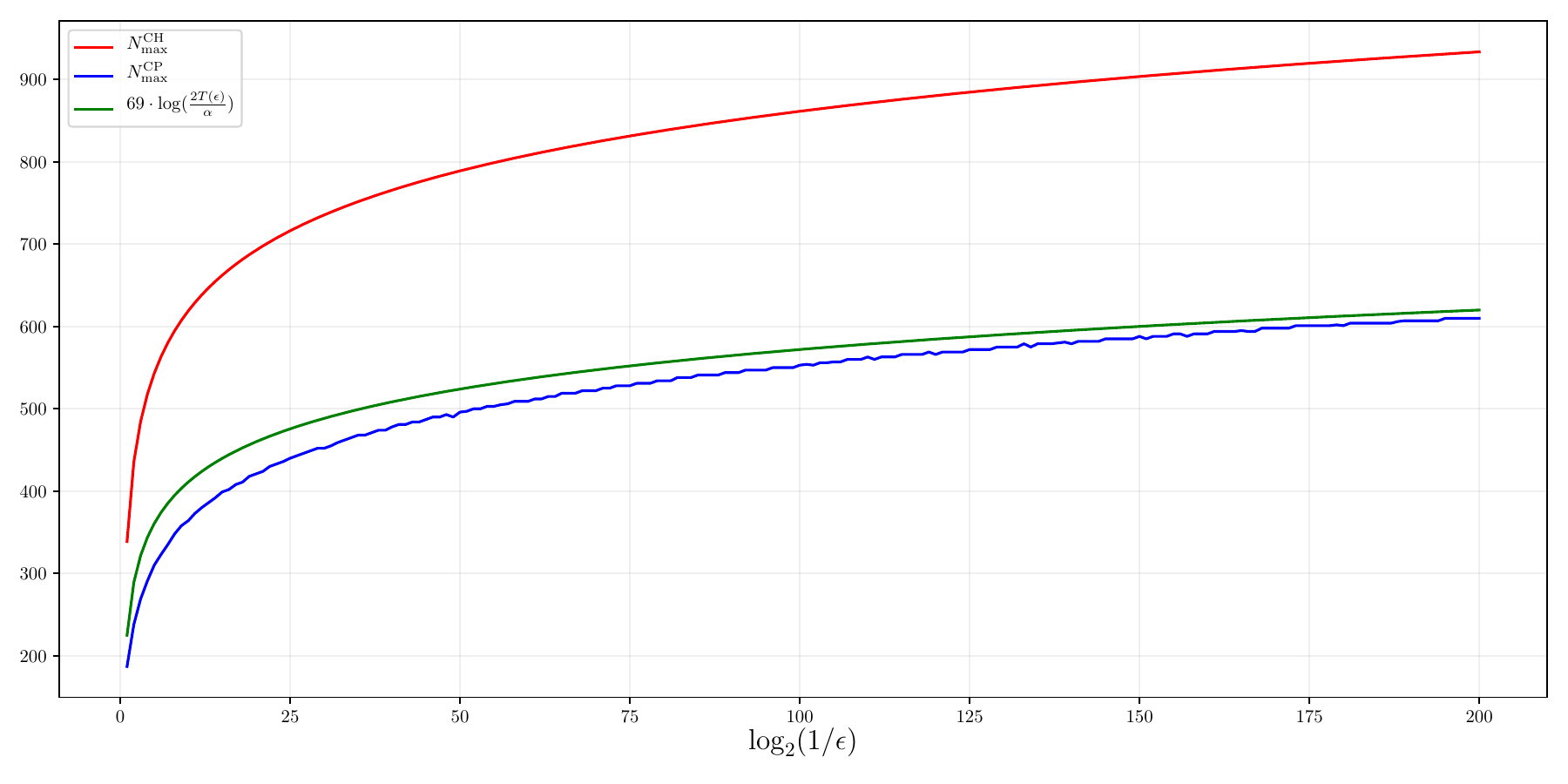}
\caption{
Functions $N^{\text{CH}}_{\text{max}} $ and $N^{\text{CP}}_{\text{max}}$ are defined by (\ref{eq:n_max_ch}) and (\ref{eq:n_max_cp}) respectively for fixed $\alpha = 0.05$. 
They represent the smallest sufficient number of shots for a round, such that (\ref{eq:qi}) is satisfied, and correspond to Chernoff-Hoeffding and Clopper-Pearson methods respectively. 
The green function (\ref{eq:n_max_cp_ch_form}) is a close approximation of the Clopper-Pearson exact blue result, which brings the bound to the same analytic form as in Chernoff-Hoeffding case.
}
\label{fig:cp_plot}
\end{figure*}

In contrast to Chernoff-Hoeffding bound, this confidence interval depends on sampled value $x$ of Binomial random variable $X$, and it is no longer possible to introduce the parameter $L$ independent of $X$ to parametrize $N_{\text{max}}$, as function of $\epsilon$ and $\alpha$, and find optimal value $L^*$ as was done in Lemma~\ref{thm:lemma}. Therefore we will find numerically the smallest sufficient number of shots $N_{\text{max}}$, such that the conditions (\ref{eq:clopper-pearson-i}) and (\ref{eq:qi}) are satisfied for each round with a given confidence level $\alpha$ and precision $\epsilon$. In order to study that, we introduce in analogy to $g_L(\theta_i)$ the function $h_{N,\epsilon,\alpha}(\theta_i) $ defined in parametric way:
\begin{eqnarray}
h_{N,\epsilon,\alpha}(\theta_i(t)) \defeq \frac{1}{2} \arccos(1-2a^{\text{min}}\left(t,N,\frac{\alpha}{T(\epsilon)}\right)) \nonumber \\
- \frac{1}{2} \arccos(1-2a^{\text{max}}\left(t,N,\frac{\alpha}{T(\epsilon)}\right)) \nonumber \\
\theta_i(t) \defeq \frac{1}{2} \arccos(1-2a^{\text{min}}\left(t,N,\frac{\alpha}{T(\epsilon)}\right)) \nonumber \\
+ \frac{1}{2} \arccos(1-2a^{\text{max}}\left(t,N,\frac{\alpha}{T(\epsilon)}\right)) \nonumber
\label{eq:h_func}
\end{eqnarray}
for $t \in \{0,1,\cdots,N\}$. In order to find optimal number of shots $N_{\text{max}}(\epsilon,\alpha)$, we need to solve the condition (\ref{eq:fmax_cond_L*}) for function $h_{N,\epsilon,\alpha}$:
\begin{eqnarray}
h_{N_{\text{max}}(\epsilon,\alpha),\epsilon,\alpha}\left(\frac{8\pi}{21}\right) = \frac{\pi}{21}
\label{eq:n_max_cp}
\end{eqnarray}
We solve this numerically for fixed $\alpha = 0.05$ and $\epsilon \geq 2^{-200}$ and upper bound the result (see Fig.~\ref{fig:cp_plot}) with 
\begin{eqnarray}
N_{\text{max}}(\epsilon, \alpha) = 69 \log(\frac{2}{\alpha} \log_2 \left( \frac{\pi}{4\epsilon} \right))
\label{eq:n_max_cp_ch_form}
\end{eqnarray}
To get the corresponding bound for $N_{\text{oracle}}$ in accordance with (\ref{eq:n_oracle_bound}) we need to calculate numerically $L_{\text{max}}(\epsilon,\alpha)$, where
\begin{equation}
L_{\text{max}}(\epsilon,\alpha) \defeq \max_{\theta} h_{N_{\text{max}}(\epsilon,\alpha),\epsilon,\alpha}(\theta).
\label{def:LmaxCP}
\end{equation}
We bound this function by a number $L^{*}_{\text{max}} > L_{\text{max}}(\epsilon,0.05)$ for every $\epsilon$ in our range. Numerically we get $L^{*}_{\text{max}} = 0.161$, which gives us the declared bound:
\begin{eqnarray}
N_{\text{oracle}} 
&<& \frac{14}{\epsilon} \log(\frac{2}{\alpha}\log_2 \left (\frac{\pi}{4\epsilon}  \right)) \nonumber .
\end{eqnarray}
\end{proof}

\bibliographystyle{IEEEtranN}
\bibliography{references}

\begin{thebibliography}{24}
\providecommand{\natexlab}[1]{#1}
\providecommand{\url}[1]{#1}
\csname url@samestyle\endcsname
\providecommand{\newblock}{\relax}
\providecommand{\bibinfo}[2]{#2}
\providecommand{\BIBentrySTDinterwordspacing}{\spaceskip=0pt\relax}
\providecommand{\BIBentryALTinterwordstretchfactor}{4}
\providecommand{\BIBentryALTinterwordspacing}{\spaceskip=\fontdimen2\font plus
\BIBentryALTinterwordstretchfactor\fontdimen3\font minus
  \fontdimen4\font\relax}
\providecommand{\BIBforeignlanguage}[2]{{%
\expandafter\ifx\csname l@#1\endcsname\relax
\typeout{** WARNING: IEEEtranN.bst: No hyphenation pattern has been}%
\typeout{** loaded for the language `#1'. Using the pattern for}%
\typeout{** the default language instead.}%
\else
\language=\csname l@#1\endcsname
\fi
#2}}
\providecommand{\BIBdecl}{\relax}
\BIBdecl

\bibitem[Brassard et~al.(2002)Brassard, Hoyer, Mosca, and Tapp]{brassard}
G.~Brassard, P.~Hoyer, M.~Mosca, and A.~Tapp, ``{Quantum Amplitude
  Amplification and Estimation},'' \emph{Contemporary Mathematics}, vol. 305,
  2002.

\bibitem[Woerner and J.~Egger(2019)]{wor}
S.~Woerner and D.~J.~Egger, ``Quantum risk analysis,'' \emph{npj Quantum
  Information}, vol.~5, 12 2019.

\bibitem[Egger et~al.(2020)Egger, Gutierrez, Mestre, and Woerner]{Egger2019}
D.~J. Egger, R.~G. Gutierrez, J.~C. Mestre, and S.~Woerner, ``Credit risk
  analysis using quantum computers,'' \emph{IEEE Transactions on Computers},
  2020.

\bibitem[Rebentrost et~al.(2018)Rebentrost, Gupt, and Bromley]{Rebentrost2018}
\BIBentryALTinterwordspacing
P.~Rebentrost, B.~Gupt, and T.~R. Bromley, ``Quantum computational finance:
  Monte carlo pricing of financial derivatives,'' \emph{Phys. Rev. A}, vol.~98,
  p. 022321, Aug 2018. [Online]. Available:
  \url{https://link.aps.org/doi/10.1103/PhysRevA.98.022321}
\BIBentrySTDinterwordspacing

\bibitem[Stamatopoulos et~al.(2020)Stamatopoulos, Egger, Sun, Zoufal, Iten,
  Shen, and Woerner]{Stamatopoulos2019}
N.~Stamatopoulos, D.~J. Egger, Y.~Sun, C.~Zoufal, R.~Iten, N.~Shen, and
  S.~Woerner, ``Option pricing using quantum computers,'' \emph{Quantum},
  vol.~4, p. 291, 2020.

\bibitem[Zoufal et~al.(2019)Zoufal, Lucchi, and Woerner]{Zoufal2019}
C.~Zoufal, A.~Lucchi, and S.~Woerner, ``Quantum generative adversarial networks
  for learning and loading random distributions,'' \emph{npj Quantum
  Information}, vol.~5, no.~1, pp. 1--9, 2019.

\bibitem[Montanaro(2015)]{Montanaro2017}
A.~Montanaro, ``Quantum speedup of monte carlo methods,'' \emph{Proceedings of
  the Royal Society A: Mathematical, Physical and Engineering Sciences}, vol.
  471, no. 2181, p. 20150301, 2015.

\bibitem[Nielsen and Chuang(2010)]{nielsen}
M.~A. Nielsen and I.~L. Chuang, \emph{Quantum Computation and Quantum
  Information}.\hskip 1em plus 0.5em minus 0.4em\relax Cambridge University
  Press, 2010.

\bibitem[Shor(1997)]{Shor1997}
\BIBentryALTinterwordspacing
P.~W. Shor, ``{Polynomial-time algorithms for prime factorization and discrete
  logarithms on a quantum computer},'' \emph{SIAM Journal on Computing},
  vol.~26, no.~5, pp. 1484--1509, aug 1997. [Online]. Available:
  \url{http://arxiv.org/abs/quant-ph/9508027}
\BIBentrySTDinterwordspacing

\bibitem[Suzuki et~al.(2020)Suzuki, Uno, Raymond, Tanaka, Onodera, and
  Yamamoto]{Suzuki2019}
Y.~Suzuki, S.~Uno, R.~Raymond, T.~Tanaka, T.~Onodera, and N.~Yamamoto,
  ``Amplitude estimation without phase estimation,'' \emph{Quantum Information
  Processing}, vol.~19, no.~2, pp. 1--17, 2020.

\bibitem[Wie(2019)]{Wie2019}
\BIBentryALTinterwordspacing
C.~R. Wie, ``{Simpler quantum counting},'' \emph{Quantum Information and
  Computation}, vol.~19, no. 11-12, pp. 967--983, jul 2019. [Online].
  Available: \url{http://arxiv.org/abs/1907.08119}
\BIBentrySTDinterwordspacing

\bibitem[Kitaev(1995)]{Kitaev1995}
\BIBentryALTinterwordspacing
A.~Y. Kitaev, ``{Quantum measurements and the Abelian Stabilizer Problem},''
  1995. [Online]. Available: \url{http://arxiv.org/abs/quant-ph/9511026}
\BIBentrySTDinterwordspacing

\bibitem[Kitaev et~al.(2002)Kitaev, Shen, and Vyalyi]{kitaev2002classical}
A.~Y. Kitaev, A.~Shen, and M.~N. Vyalyi, \emph{Classical and quantum
  computation}.\hskip 1em plus 0.5em minus 0.4em\relax American Mathematical
  Soc., 2002, no.~47.

\bibitem[Svore et~al.(2014)Svore, Hastings, and Freedman]{svore2013faster}
K.~M. Svore, M.~B. Hastings, and M.~Freedman, ``Faster phase estimation,''
  \emph{Quantum Information \& Computation}, vol.~14, no. 3-4, pp. 306--328,
  2014.

\bibitem[Atia and Aharonov(2017)]{atia2017fast}
Y.~Atia and D.~Aharonov, ``Fast-forwarding of hamiltonians and exponentially
  precise measurements,'' \emph{Nature communications}, vol.~8, no.~1, pp.
  1--9, 2017.

\bibitem[Aaronson and Rall(2020)]{Aaronson2019}
S.~Aaronson and P.~Rall, ``Quantum approximate counting, simplified,'' in
  \emph{Symposium on Simplicity in Algorithms}.\hskip 1em plus 0.5em minus
  0.4em\relax SIAM, 2020, pp. 24--32.

\bibitem[Koch(1999)]{koch_1999_parameter}
K.-R. Koch, \emph{Parameter Estimation and Hypothesis Testing in Linear
  Models}.\hskip 1em plus 0.5em minus 0.4em\relax Springer-Verlag Berlin
  Heidelberg, 1999.

\bibitem[Clopper and Pearson(1934)]{ClopperPearson1934}
C.~Clopper and E.~Pearson, ``The use of confidence or fiducial limits
  illustrated in the case of the binomial,'' \emph{Biometrika}, vol.~26, no.~4,
  pp. 404--413, dec 1934.

\bibitem[Scholz(2008)]{scholz2008confidence}
F.~Scholz, ``Confidence bounds and intervals for parameters relating to the
  binomial negative binomial poisson and hypergeometric distributions with
  applications to rare events,'' 2008.

\bibitem[Hoeffding(1963)]{hoeffding1963probability}
W.~Hoeffding, ``Probability inequalities for sums of bounded random
  variables,'' \emph{Journal of the American Statistical Association}, pp.
  13--30, 1963.

\bibitem[{A. H{\'e}ctor et al.}(2019)]{qiskit}
{A. H{\'e}ctor et al.}, ``Qiskit: An open-source framework for quantum
  computing,'' 2019.

\bibitem[Burchard(2019)]{burchard2019lower}
P.~Burchard, ``Lower bounds for parallel quantum counting,'' \emph{arXiv
  preprint arXiv:1910.04555}, 2019.

\bibitem[Maldonado and Greenland(1994)]{Maldonado1994}
\BIBentryALTinterwordspacing
G.~Maldonado and S.~Greenland, ``A comparison of the performance of model-based
  confidence intervals when the correct model form is unknown,''
  \emph{Epidemiology}, vol.~5, no.~2, pp. 171--182, Mar. 1994. [Online].
  Available: \url{https://doi.org/10.1097/00001648-199403000-00007}
\BIBentrySTDinterwordspacing

\bibitem[Jeng and Meeker(2000)]{Jeng2000}
\BIBentryALTinterwordspacing
S.-L. Jeng and W.~Q. Meeker, ``Comparisons of approximate confidence interval
  procedures for type i censored data,'' \emph{Technometrics}, vol.~42, no.~2,
  pp. 135--148, 2000. [Online]. Available:
  \url{http://www.jstor.org/stable/1271445}
\BIBentrySTDinterwordspacing

\end{thebibliography}
\end{document}